\documentclass[lettersize,journal]{IEEEtran}
\usepackage{amsmath,amsfonts}
\usepackage{algorithmic}
\usepackage{algorithm}
\usepackage{array}
\usepackage{textcomp}
\usepackage{stfloats}
\usepackage{url}
\usepackage{verbatim}
\usepackage{graphicx,subfigure}
\usepackage{amssymb}
\usepackage{cite}
\usepackage{amsthm} % proof
\usepackage{color}
\usepackage{diagbox,multirow,makecell}

\definecolor{mygreen}{rgb}{0, 0.63, 0.32}
\definecolor{mybrown}{rgb}{0.7, 0.3, 0.0}

\makeatletter % <=======================================================
\let\myorg@bibitem\bibitem
\def\bibitem#1#2\par{%
  \@ifundefined{bibitem@#1}{%
    \myorg@bibitem{#1}#2\par
  }{%
    \begingroup
      \color{\csname bibitem@#1\endcsname}%
      \myorg@bibitem{#1}#2\par
    \endgroup
  }%
}
\makeatother %

\begin{document}

\title{Joint Beamforming and Position Optimization for Fluid RIS-aided ISAC Systems}

\author{Junjie~Ye, \IEEEmembership{Graduate Student Member,~IEEE,} Peichang~Zhang, \IEEEmembership{Member,~IEEE,} Xiao-Peng~Li, \IEEEmembership{Member,~IEEE},    \\ Lei~Huang, \IEEEmembership{Senior Member,~IEEE,}  Yuanwei~Liu, \IEEEmembership{Fellow,~IEEE}
\thanks{This work has been submitted to the IEEE for possible publication. Copyright may be transferred without notice, after which this version may no longer be accessible.}
\thanks{J. Ye, X.-P. Li, P. Zhang and L. Huang are with State Key Laboratory of Radio Frequency Heterogeneous Integration, Shenzhen University, Shenzhen, China. (e-mail: 2152432003@email.szu.edu.cn; $\lbrace$pzhang, x.p.li, lhuang$\rbrace$@szu.edu.cn.) \\ %(Corresponding author: L. Huang; P. Zhang.)
\indent Y. Liu is with the Department of Electrical and Electronic Engineering, the
University of Hong Kong, Hong Kong, China. (e-mail: yuanwei@hku.hk).}
}
%\thanks{This work is supported in part by the Key Project of International Cooperation and Exchanges of the National Natural Science Foundation of China under Grant 62220106009, the National Science Fund for Distinguished Young Scholars under Grant 61925108,  the project of Shenzhen Peacock Plan Teams under Grant KQTD20210811090051046 and Research Team Cultivation Program of Shenzhen University under Grant 2023DFT003.}

\maketitle

\begin{abstract}
A fluid reconfigurable intelligent surface (fRIS)-aided integrated sensing and communication (ISAC) system is proposed to enhance multi-target sensing and multi-user communication. Unlike the conventional RIS, the fRIS employs movable elements with adjustable positions, offering additional spatial degrees of freedom. In this system, a joint optimization problem is formulated to minimize sensing beampattern mismatch and symbol estimation error. {An algorithm based on alternating minimization is devised to handle the resultant non-convex problem,}  where the subproblems are solved via augmented Lagrangian method, quadratic programming, semidefinite relaxation, and majorization-minimization. A key challenge is that the element positions affect both incident and reflective channels, leading to the high-order composite objective functions. As a remedy, the high-order terms are transformed into linear and linear-difference forms by exploiting the structural characteristics of fRIS and the channels. Numerical results demonstrate the superiority of the proposed scheme over conventional RIS-aided ISAC and other benchmarks.
\end{abstract}

\begin{IEEEkeywords}
Alternating minimization, fluid reconfigurable intelligent surface, integrated sensing and communications, mismatch and estimation error
\end{IEEEkeywords}

\section{Introduction}
\IEEEPARstart{I}{n} the era of sixth-generation wireless communications, integrated sensing and communications (ISAC) have emerged as a promising paradigm to alleviate resource scarcity \cite{ISAC_Survey}. In ISAC, the communication system and the sensing system aim to share the same spectrum \cite{JRC_coexistence}, where various schemes have been developed, such as null-space projection \cite{NSP} and transceiver design \cite{Rihan_RadCom}. Furthermore, the works \cite{Com_Sensing_collaborate, Com_Sensing_cooperate} achieved collaborations of the two systems by sharing critical information. Additionally, given the hardware and signal processing similarities between sensing and communication, recent ISAC research focused on unified platforms capable of both functions \cite{El3}. For instance, the work \cite{Joint_Transmit_Beamforming} designed the transmit beamformer by combining radar waveforms and communication symbols, while the work \cite{ZJ_OFDM_ISAC} addressed wideband ISAC using a unified waveform design metric. In \cite{Qihao2}, a latency-aware resource allocation scheme was developed for multiple-input and multiple-output ISAC systems.
{In addition, recent works explored parameter estimation in ISAC systems. For example, the work \cite{ISACEstimate} proposed a hybrid fusion framework, where the fused signal was separated for the estimation of the target directions. }

Meanwhile, reconfigurable intelligent surfaces (RIS) have attracted great attention. Comprised of many reflective elements, RIS is capable of inducing phase shifts on incident signals and dynamically reconfiguring wireless channels \cite{CunhuaRISSignalProcess}. This capability enables substantial communication sum-rate gains in multi-user scenarios \cite{Sumrate_RIS}, and facilitates suppression of undesired signals such as eavesdropping \cite{RIS_eavesdrop} and jamming \cite{RIS_suppress}. Besides,  the work \cite{Qihao} showed that RIS can construct favorable channels to support devices with ultra-reliable and low-latency communications. In addition, RIS can be applied to various areas, such as radar detection \cite{Rihan_RIS_radar}, near-field communications \cite{LYW_nearfield}, and wireless power transfer and communication \cite{FJ_power}.

To harness the combined benefits of ISAC and RIS, extensive studies have been conducted on RIS-aided ISAC systems \cite{Big_Survey}. In \cite{JZM_DFRC}, the authors maximized the echo signal-to-noise ratio (SNR) by jointly designing the transmitter and RIS beamformer in a single-user and single-target scenario. This framework was extended in \cite{RIS_ISAC} to support multi-user and multi-target systems. {Furthermore, the work \cite{LDRRISISAC} addressed the issue of high dynamic range caused by strong interference in RIS-aided ISAC systems, where a joint active and passive beamforming was designed to reduce the dynamic range and mitigate the interference.} To reduce energy consumption, energy efficiency optimization was explored in RIS-aided ISAC systems \cite{PassiveRISDFRCTGCN, YJJ_EE}. Additionally, different applications were investigated, including simultaneously transmitting and reflecting surfaces \cite{Star}, unsupervised learning \cite{Unsupervised}, and extended reality \cite{XR_RIS_ISAC}.

In the above works, there remains potential to improve spatial degrees of freedom (DoFs) and reduce system complexity. Particularly, the conventional arrays have fixed antennas after assembly, which restricts the spatial diversity \cite{FA_Capacity}. Furthermore, achieving high performance often requires a large number of elements, leading to increased manufacturing costs and optimization burdens. To overcome these issues, the fluid antenna (FA), also known as movable antenna, has emerged as a candidate. Enabled by fluidic metamaterials, FA can dynamically adjust the antenna positions, offering additional spatial DoFs \cite{ FA_basic2}. Moreover, the improved performance enables comparable gains with fewer elements, thereby reducing the computational load of large-scale optimization. The potential of FA was extensively explored, such as point-to-point capacity  \cite{FA_Capacity}, communication sum-rate \cite{FA_MU}, and ISAC performance improvement  \cite{FA_ISAC, FA_ISAC2, FA_ISAC3}. 

To adaptively control the propagation environment, some efforts incorporated RIS in FA systems \cite{RIS_FA, fasriscomm}. However, they still rely on fixed RIS elements. Considering that RIS has more elements, fluid RIS (fRIS) or movable-element RIS  was studied \cite{fRISone,fRIStwo,fRISthree,MERISone,MERIStwo,MERISthree,liuyu} to exploit more DoFs, where the elements in RIS are movable. {Compared to the RIS-aided FAS system, fRIS has several advantages. First, fRIS can bring more spatial DoFs, as the RIS typically contains more elements than the BS antennas. Second, RIS elements are passive and free from the RF chain constraints. Third, from a deployment perspective, fRISs are typically more ubiquitous than FAS-BSs in large-scale networks, effectively improving system performance.}

{In the fRIS systems, several works \cite{fRISone,fRIStwo,fRISthree}  explored pixel-based hardware architectures that emulate virtual element movement through densely distributed port switching.  In contrast, other works  \cite{MERISone,MERIStwo,MERISthree,liuyu} employed flexible cables, motors, and sliding tracks to reposition the elements for communication enhancement. {While these implementations inevitably add hardware complexity, the fRIS can achieve improved performance with fewer elements due to the additional DoFs in a new dimension\footnote{{The fRIS introduces new spatial DoFs by allowing controlled displacement of each reflecting element as compared to the traditional RIS. These position-based DoFs are fundamentally distinct from phase-shift control, as they reshape the propagation geometry, offering a complementary optimization domain. }}. Additionally, enabling technologies are actively studied and continuously refined, making the hardware complexity manageable. This offers a new direction for RIS evolution.} However, these works concerning the fRIS mainly focus on basic communication settings and lack  ISAC modeling and optimization, motivating further study in this direction.}

Against the above background, we investigate a fRIS-aided ISAC system, where the fRIS forms beams to simultaneously serve communication users and illuminate sensing targets. A joint optimization problem is formulated to minimize both sensing beampattern mismatch and communication decoding error. Then, the resultant problem is addressed using the alternating minimization (AM)-based algorithm incorporating augmented Lagrangian method (ALM), quadratic programming (quadprog), semidefinite relaxation (SDR), and majorization-minimization (MM). The main contributions are summarized as follows:
\begin{itemize}
    \item  Leveraging the additional spatial DoFs introduced by movable elements, we propose a novel fRIS-aided ISAC system to enhance sensing and communication. Specifically, we characterize the transmit signal, fRIS beampattern, symbol decoding, and fRIS-related channels. Besides, we formulate a joint minimization problem for sensing mismatch and decoding error.
    
    \item  To solve the resultant problem, we devise an AM-based algorithm. First, the original high-order formulation is transformed into a more tractable form. Then, the problem is decomposed into subproblems, which are tackled using dedicated optimization techniques. Furthermore, we analyze the convergence and computational complexity of the proposed algorithm.
    
    \item  The subproblem of optimizing fRIS element positions is particularly challenging, as their movement affects both incident and reflective channels. To address this issue, we exploit the structural properties of fRIS and channels to reduce the fourth-order terms to linear and linear-difference forms, simplifying the objective and facilitating the optimization.
    
    \item  Numerical results validate the effectiveness of the proposed model, showing notable performance gains over existing benchmarks. Moreover, fRIS compensates for performance degradation caused by reducing the number of RIS elements, alleviating the high-dimensional complexity in RIS optimization.
\end{itemize}

This paper is organized as follows. Section \ref{sec:model} presents the architecture of the fRIS-aided ISAC system and formulates the joint minimization problem. Section \ref{algorithm} details the proposed AM-based algorithm, while analyzes its convergence and complexity. Numerical results are shown in Section \ref{simulation}, and conclusions are drawn in Section \ref{conclusions}.

\textit{Notation}: Bold lowercase and uppercase letters denote vectors and matrices, respectively. In addition,  $(\cdot)^\mathrm{H}$, $(\cdot)^\mathrm{T}$, and $(\cdot)^*$ signify conjugate transpose, transpose, and conjugate, respectively. $\mathrm{tr}(\cdot)$ is the trace of a matrix. $\mathrm{diag}(\cdot)$ extracts diagonal entries into a vector, while $\mathrm{Diag}(\cdot)$ creates a diagonal matrix from a vector. $\mathrm{vec}(\mathbf{A})$ vectorizes matrix $\mathbf{A}$. $|\cdot|$ and $\|\cdot\|$ denote absolute value and norm, respectively. $\mathfrak{R}[x]$ and $\angle x$ are the real part and phase of a complex scalar $x$, respectively.

\section{System Model and Problem Formulation} \label{sec:model}
\subsection{System Model}
\begin{figure}[!t]
    \centering
    \includegraphics[scale=0.35]{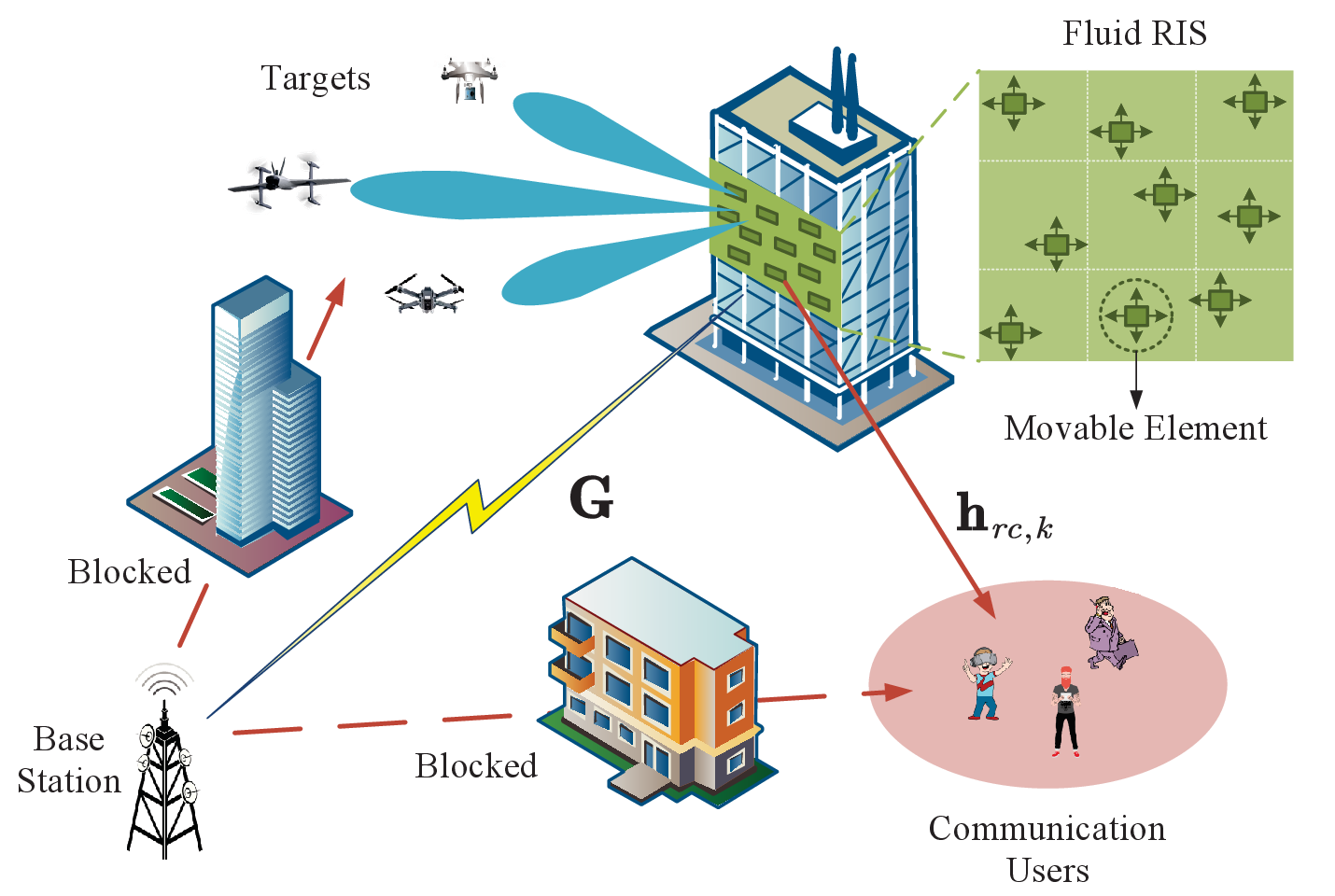}
    \caption{A scenario of a fRIS-aided ISAC system that performs multi-target sensing and multi-user communication.}
    \label{fig:System model}
\end{figure}

As illustrated in Fig. \ref{fig:System model}, we consider an ISAC system assisted by a fRIS. The ISAC base station (BS), equipped with a uniform linear array (ULA) of $M$ antennas, simultaneously serves $K$ single-antenna communication users and senses $T$ point targets. Without loss of generality, we assume that direct links for both sensing and communication are blocked and thus a fRIS is deployed to establish virtual communication and sensing paths.  The fRIS is comprised of $N$ movable elements, where each element can reposition freely within a planar region $\mathcal{A}$ of size $A \times A$. The position of the $n$-th element is denoted as $\mathbf{p}_n = [p_{x,n}, p_{y,n}]^\mathrm{T} \in \mathcal{A}^{2 \times 1}$, where all the element positions are given by $\mathbf{p} = [\mathbf{p}_1, \mathbf{p}_2, \ldots, \mathbf{p}_N] \in \mathcal{A}^{2 \times N}$. %\textcolor{blue}{Besides, it is assumed that the control latency of element mobility during reconfiguration is negligible\footnote{\textcolor{mybrown}{In mechanically actuated designs, the latency mainly stems from the motors, where the delay is acceptable under quasi-static channel conditions. In pixel-based schemes, port switching is near-instantaneous due to electronic actuation. In dynamic environments, predictive schemes (e.g., channel prediction) may help initiate early element movement.}}. }

In this system, the signal $\mathbf{x} \in \mathbb{C}^{M}$ is transmitted from the BS, which acts as dual-functional waveform
for both sensing and communication. The transmit signal propagates through the channel $\mathbf{G} \in \mathbb{C}^{N \times M} $ to the fRIS. In the fRIS, the phase of the incident signal $\mathbf{G} \mathbf{x}$ is altered by an adjustable phase shift matrix $\mathbf{\Theta} = \mathrm{Diag}\left(\theta_1, \theta_2, \cdots, \theta_N\right)$, in which $ \theta_n$ represents the phase shift coefficient for the $n$-th fRIS element. As a result, the reflective signal at the fRIS is given by
\begin{align}
    \mathbf{v} = \boldsymbol{\Theta}^\mathrm{H} \mathbf{G} \mathbf{x}.
\end{align}
Subsequently, the reflective signal at the fRIS ought to form multiple beams to illuminate the targets and serve the communication users via the reflecting channel $\mathbf{H}_{rc}$, where $\mathbf{H}_{rc} =\begin{bmatrix}
        \mathbf{h}_{rc,1} & \!\!\!
        \cdots \!\!\!& \mathbf{h}_{rc,k} & \!\!\!
        \cdots \!\!\!&
        \mathbf{h}_{rc,K}
    \end{bmatrix}\in \mathbb{C}^{N \times K}$ represents the overall downlink reflective channel while $\mathbf{h}_{rc,k}^\mathrm{H}\in \mathbb{C}^{1 \times N}$ is the reflective channel from the fRIS to the $k$-th user. 

\subsubsection{Sensing Model}
In the fRIS, beams should be formed towards the directions of the targets for target illuminations. To this end, we can design $\mathbf{v}$ to manipulate the beampattern of the reflective signal. Specifically, the beampattern  of $\mathbf{v}$ is expressed as
   \begin{align} \label{beampattern_orig}
    \mathcal{P}_s(\phi,\psi,\mathbf{p}, \mathbf{R}_s) &= \mathbf{a}^\mathrm{H}(\phi,\psi,\mathbf{p})\mathbf{R}_s\mathbf{a}(\phi,\psi,\mathbf{p}),
\end{align} 
which indicates the illumination power of the fRIS at the azimuth angle $\phi$ and the elevation angle $\psi$. In (\ref{beampattern_orig}), we define $\mathbf{R}_s =\mathbf{v}\mathbf{v}^\mathrm{H}=\boldsymbol{\Theta}^\mathrm{H} \mathbf{G} \mathbf{x} \mathbf{x}^\mathrm{H} \mathbf{G}^\mathrm{H} \boldsymbol{\Theta}$, while the symbol of $\mathbf{a}(\phi,\psi,\mathbf{p})$ represents the steering vector of the fRIS. For conciseness, we present the detailed expression of $\mathbf{a}(\phi,\psi,\mathbf{p})$  in Subsection \ref{subsubsubsec: channel}.

For beam design, a common metric is the beampattern similarity, which evaluates the mismatch between $\mathcal{P}_s(\phi,\psi,\mathbf{p}, \mathbf{R}_s)$ and an ideal pattern $\mathcal{P}_d(\phi,\psi)$. The mismatch is quantified as
\begin{align} \label{sensing_loss}
    \varepsilon_r = \sum^{I_a}_{i=1}\sum^{I_e}_{i'=1} \left|\beta \mathcal{P}_d(\phi_i,\psi_{i'}) -\mathcal{P}_s(\phi_i,\psi_{i'})\right|^2,
\end{align}
where $\beta$ is a scaling factor, while $I_a$ and $I_e$ denote azimuth and elevation angle sample numbers, respectively. Minimizing $\varepsilon_r$ encourages $\mathcal{P}_s(\phi,\psi,\mathbf{p}, \mathbf{R}_s)$ to closely match $\mathcal{P}_d(\phi_i,\psi_{i'})$, enhancing the target illumination performance.

\subsubsection{Communication Model}
After $\mathbf{v}$ propagates to the communication users, the overall received signal $\mathbf{y}_{c} \in \mathbb{C}^{K}$ at the users is given by
\begin{align} \label{com_receive}
    \mathbf{y}_{c} = \mathbf{H}_{rc}^\mathrm{H} \boldsymbol{\Theta}^\mathrm{H} \mathbf{G} \mathbf{x} + \mathbf{n}_{c}, 
\end{align}
where $\mathbf{n}_{c} \in \mathbb{C}^{K}$ represents the additive Gaussian white noise matrix, whose entries are independent and identically distributed with zero mean and $\sigma_0^2$ variance. We denote $y_{c,k}$ as the received signal of the $k$-th user, which is the $k$-th entry of $\mathbf{y}_{c}$.   Given a desired constellation symbol ${{s}}_{c,k}$ for the $k$-th user, an estimation factor $\omega$ is introduced to recover ${{s}}_{c,k}$ from $y_{c,k}$, which is given by
\begin{align} \label{com_estimation}
    \hat{s}_{c,k} = \omega y_{c,k},
\end{align}
where $\hat{s}_{c,k}$ is the estimated symbol. To evaluate the communication performance, a direct metric is the mean square error (MSE) between the estimated symbol and the transmitted symbol: %  \cite{comMSE,comMSE2,comMSE3,liaobin_kkt}
  \begin{align} \label{comm_loss}
    \varepsilon_c &= \mathbb{E}[\|\mathbf{s}_c -  \hat{\mathbf{s}}_c\|^2] \notag \\ 
    &{=} \|\mathbf{s}_c -  \omega\mathbf{H}_{rc}^{\mathrm{H}} \boldsymbol{\Theta}^\mathrm{H} \mathbf{G} \mathbf{x}\|^2 + K \omega^2 \sigma_0^2,
\end{align}   
where $\mathbf{s}_c \in \mathbb{C}^{K}$ and $\hat{\mathbf{s}}_c\in \mathbb{C}^{K}$ incorporate all the ${{s}}_{c,k}$ and $\hat{s}_{c,k}$, respectively. In (\ref{comm_loss}), the first term can be viewed as the multi-user interference, while the second term is the impact of the noise.

\subsubsection{Channel Model} \label{subsubsubsec: channel}
The steering vector of the fRIS is dependent on the positions of the elements $\mathbf{p}$, the azimuth angle $\phi$, and the elevation angle $\psi$. Specifically, the steering vector of the fRIS is expressed as \cite{FA_Capacity}: 
\begin{align} \label{FA_steering_vector}
    \!\!\mathbf{a}(\phi,\psi,\mathbf{p})\!=\!\begin{bmatrix}
            e^{\jmath \frac{2\pi}{\lambda}d_{1}(\phi,\psi,\mathbf{p}_1) }  & \!\! \!\cdots\!\!\! & e^{\jmath\frac{2\pi}{\lambda} d_{N}(\phi,\psi,\mathbf{p}_N)} 
        \end{bmatrix}^\mathrm{T}\!\!\!,
\end{align}
 where $d_{n}(\phi,\psi, \mathbf{p})$ is referred to as the path difference of the signal propagation from the direction $(\phi,\psi)$ between the $n$-th element and the reference origin, given by 
    \begin{align} \label{path_diff}
        d_{n}(\phi,\psi,\mathbf{p}_n) = p_{x,n} \sin (\phi)\cos (\psi) + p_{y,n} \sin (\psi).
    \end{align}

Without loss of generality, the fRIS is considered deployed in the area where line-of-sight (LoS) paths dominate, such that the signal coverage can be extended. Consequently, the fRIS-related channels are modeled as LoS channels. The channel from the BS to the fRIS is modeled as
\begin{align} \label{chan_G}
    \mathbf{G}& =\sqrt{\zeta_G} \mathbf{a}_r(\phi_r, \psi_r, \mathbf{p})  \mathbf{a}_t^{\mathrm{H}}(\phi_t),
\end{align}
in which $\sqrt{\zeta_G}$ denotes the path loss between the BS and the fRIS, while $\mathbf{a}_r(\phi_r, \psi_r, \mathbf{p})$ represents the receive steering vector of the fRIS with the signal coming from $\phi_r$ in azimuth and $\psi_r$ in elevation. On the other hand, as the ULA is equipped in the BS, the transmit steering vector in the direction of $\phi_t$ is written as
\begin{align} \label{chan_hrc}
    \mathbf{a}_t = \begin{bmatrix}
            e^{\jmath\frac{2\pi }{\lambda} \widetilde{d}_1  }  & \cdots &e^{\jmath\frac{2\pi }{\lambda} \widetilde{d}_m  }& \cdots & e^{\jmath \frac{2\pi }{\lambda} \widetilde{d}_{M-1} } 
        \end{bmatrix}^\mathrm{T},
\end{align}
where $\lbrace\widetilde{d}_m= \frac{\lambda}{2} m\sin(\phi_t), ~\forall m=0,\cdots,M-1 \rbrace$ and the inter-element spacing is half wavelength. Similarly, the channel  from the fRIS to the $k$-th user is expressed as
\begin{align}
    \mathbf{h}_{rc,k}^\mathrm{H} =  \sqrt{\zeta_k}\mathbf{a}^\mathrm{H}_{c,k}(\phi_{c,k}, \psi_{c,k}, \mathbf{p}), \forall k = 1,\cdots, K, 
\end{align}
where $\mathbf{a}_{c,k}(\phi_{c,k}, \psi_{c,k}, \mathbf{p})$ denotes the steering vector of the fRIS to the $k$-th user from $\phi_{c,k}$ and $\psi_{c,k}$. The path loss between the fRIS and the $k$-th user is denoted as $\sqrt{\zeta_k}$. Accordingly, the overall downlink reflective channel $\mathbf{H}_{rc}$ is rewritten as
\begin{align}
    \mathbf{H}_{rc} &= \begin{bmatrix}
        \mathbf{h}_{rc,1} & \!\!\!
        \cdots \!\!\!&
        \mathbf{h}_{rc,K}
    \end{bmatrix} = \mathbf{A}_{rc} \boldsymbol{\Sigma}_{rc},
\end{align}
where $\mathbf{A}_{rc}$ and $\mathbf{\Sigma}_{rc}$ are defined as 
\begin{subequations} \label{chan_Hrc_aff}
  \begin{align}
\mathbf{A}_{rc}&=\begin{bmatrix}
        \mathbf{a}_{c,1}(\phi_{c,1}, \psi_{c,1}, \mathbf{p}) & \!\!\!
        \cdots \!\!\!&
        \mathbf{a}_{c,K}(\phi_{c,K}, \psi_{c,K}, \mathbf{p})
    \end{bmatrix},\\
    \boldsymbol{\Sigma}_{rc} &= \mathrm{Diag}\left(\begin{bmatrix}        \sqrt{\zeta_1}&\cdots&\sqrt{\zeta_K}
    \end{bmatrix}\right).
\end{align}  
\end{subequations}
\textit{{Note: In practice, RIS element positions may deviate slightly from their designed values due to hardware imperfections. Therefore, deviation $\mathbf{p}=\overline{\mathbf{p}}+\Delta \mathbf{p}$ may introduce phase shifts offset in the steering vector $\mathbf{a}(\mathbf{p})=\boldsymbol{\Pi}(\Delta\mathbf{p}) ~\overline{\mathbf{a}}(\overline{\mathbf{p}})$. It can be analyzed that the effect on communication is negligible due to phase cancellation, while the sensing performance may experience distortion in the beampattern shape, which remains limited under small deviations. While the extension to a robust framework regarding position deviations is non-trivial, it is left to future work.}}
    
\subsection{Problem Formulation} 
To simultaneously achieve sensing and communication functions, a weighted sum of the sensing beampattern mismatch $\varepsilon_r$ of (\ref{sensing_loss}) and the communication symbol estimation MSE  $\varepsilon_c$ of (\ref{comm_loss}) is minimized by optimizing the transmit waveform $\mathbf{x}$, the fRIS phase shift $\boldsymbol{\Theta}$, the fRIS element positions $\mathbf{p}$, the communication symbol estimator $\omega$ and the scaling factor $\beta$ of the sensing beampattern. Consequently, the problem is formulated as
\begin{subequations}\label{formulation}
\begin{align} 
\min _{\mathbf{x},\boldsymbol{\Theta},\mathbf{p}, \omega, \beta } &~  \varepsilon_0\!=\!\alpha \left(\sum^{I_a}_{i=1}\sum^{I_e}_{i'=1} \left|\beta \mathcal{P}_d(\phi_i,\psi_{i'}) -\mathcal{P}_s(\phi_i,\psi_{i'})\right|^2\right) \notag\\
&\!\!+ \!(1\!- \!\alpha) \!\left(\!\|\mathbf{s}_c -  \omega\mathbf{H}_{rc}^{\mathrm{H}} \boldsymbol{\Theta}^\mathrm{H} \mathbf{G} \mathbf{x}\|^2 \!+\! K \omega^2 \sigma_0^2\!\right) \tag{\ref{formulation}{a}} \\ 
\mathrm{s.t.} &~  \|\mathbf{x}\|^2 \leq P_t, \tag{\ref{formulation}{b}}\\
&~ \omega \in \mathbb{R}, \tag{\ref{formulation}{c}} \\
&~ |\boldsymbol{\Theta}|_{n,n}=1, \forall n =1,\cdots, N  \tag{\ref{formulation}{d}} \\
&~ \mathbf{p}_n \in \mathcal{A}, \forall n =1,\cdots, N \tag{\ref{formulation}{e}} \\
&~ \|\mathbf{p}_n \!-\! \mathbf{p}_{n'}\|_2 \! \geq \! \Delta D,   \forall n,n' \!\!=\!\! 1,\!\cdots\!,N, n\!\neq \! n' , \tag{\ref{formulation}{f}} 
\end{align}
\end{subequations}
where $\alpha$ is the weighting coefficient, controlling the trade-off between sensing and  communication. The constraint of (\ref{formulation}{b}) indicates that the transmit power should not exceed $P_t$, while (\ref{formulation}{d}) requires the modulus-one phase shifts of the fRIS. {The constraints of (\ref{formulation}{e}) and (\ref{formulation}{f}) are introduced to model the physical limitations of the fRIS structure. Specifically, (\ref{formulation}{e}) captures the physical boundary of the movement area, ensuring that all elements are confined in $\mathcal{A}$.} 
{Besides, (\ref{formulation}{f}) enforces a minimum distance $\Delta D$ among elements, mitigating the electromagnetic mutual coupling effects and avoiding element collision}.

{The problem (\ref{formulation}) is non-convex owing to the two reasons. On the one hand,  (\ref{formulation}{d}) and  (\ref{formulation}{f}) are non-convex. Besides, the objective function is complex, where the term of beampattern mismatch $\varepsilon_r$ in the objective function is high-order with respect to (w.r.t.) different variables, while the variables are coupled and not jointly convex.}

\section{Proposed Algorithm} \label{algorithm}
In this section, an AM-based algorithm is developed to address (\ref{formulation}). {Given the non-convexity and NP-hardness of the problem, we solve it by iteratively updating the variables.} Additionally, we analyze the convergence and the computational complexity of the proposed algorithm.

\subsection{Reformulation of $\varepsilon_r$}\label{sec:reformulate}
{Carefully inspecting $\varepsilon_r$ in (\ref{sensing_loss}), it is a quartic function of the optimization variables, as it is quadratic w.r.t. $\mathbf{R}_s$, which itself depends quadratically on the optimization variables. } To simplify the objective function in (\ref{formulation}{a}), we reduce the order of $\varepsilon_r$ to be a quadratic function w.r.t. the variables.  First, we design a reference covariance matrix $\widetilde{\mathbf{R}}_s$ that only considers sensing. According to \cite{sensing_only}, the problem of designing $\widetilde{\mathbf{R}}_s$ can be given by
\begin{subequations}\label{formulation0}
\begin{align} 
\min _{\beta, \widetilde{\mathbf{R}}_s} &~   \sum^{I_a}_{i=1}\sum^{I_e}_{i'=1} \left|\beta \mathcal{P}_d(\phi_i,\psi_{i'}) -\mathcal{P}_s(\phi_i,\psi_{i'},\widetilde{\mathbf{R}}_s)\right|^2  \tag{\ref{formulation0}{a}} \\ 
\mathrm{s.t.} &~ \beta>0 , \widetilde{\mathbf{R}}_s \succeq 0,  \tag{\ref{formulation0}{b}} \\
& \mathrm{tr}\left(\widetilde{\mathbf{R}}_s\right)={\|\mathbf{G}\mathbf{x}\|^2},\tag{\ref{formulation0}{c}}\\
& \mathrm{rank}\left(\widetilde{\mathbf{R}}_s\right)=1, \tag{\ref{formulation0}{d}}
\end{align}
\end{subequations}
where the objective function requires the designed sensing beampattern to be similar to an ideal one. In the constraint of (\ref{formulation0}{c}), the radiation power at the fRIS is $\|\mathbf{Gx}\|^2$, and the last constraint requires $\widetilde{\mathbf{R}}_s$ to be rank-one since $\mathbf{R}_s = \mathbf{v} \mathbf{v}^\mathrm{H}$. This problem can be directly solved via SDR with Gaussian randomization \cite{SDR} and penalty-based method \cite{rank_penalty}. 

{\textit{\textbf{Remark 1:} To handle (\ref{formulation0}{d}), SDR followed by Gaussian randomization can be adopted to construct a feasible rank-one solution. While a theoretical proof of tightness is generally intractable, this approach is widely adopted and has demonstrated strong empirical performance. For the penalty-based method, the rank-one constraint can be enforced via a penalty term $\big(\mathrm{tr}(\widetilde{\mathbf{R}}_s) - \|\widetilde{\mathbf{R}}_s\|_2\big)$. As proven in \cite{penaltytwo}, a rank-one solution can be obtained when the penalty factor is sufficiently large.}}

With the obtained $\widetilde{\mathbf{R}}_s$ in (\ref{formulation0}),  the beams can be effectively formed towards the targets. Since $\widetilde{\mathbf{R}}_s$ is rank-one, a reference signal $\mathbf{s}_r$ at the fRIS associated with $\widetilde{\mathbf{R}}_s$ can be obtained by the eigenvalue decomposition of $\widetilde{\mathbf{R}}_s$, yielding
\begin{align} \label{obtain_sr}
\widetilde{\mathbf{R}}_s = \mathbf{s}_r\mathbf{s}_r^\mathrm{H},
\end{align}
where $\mathbf{s}_r = \sqrt{\lambda_s} \mathbf{u}_s$ with $\lambda_s$ and $\mathbf{u}_s$ being the maximum eigenvalue and the associated eigenvector, respectively. Here, $\mathbf{s}_r$ represents a desired and achievable reflective signal, which produces directive beams towards the targets. 

Consequently, the sensing metric is reformulated as the mismatching between $\mathbf{v}$ and $\mathbf{s}_r$, i.e.,
\begin{align}
	\widetilde{\varepsilon}_r = \|\mathbf{s}_r-\mathbf{v}\|^2 =\|\mathbf{s}_r-\boldsymbol{\Theta}^\mathrm{H} \mathbf{G} \mathbf{x}\|^2.
\end{align}
It is observed that  $\widetilde{\varepsilon}_r$ becomes quadratic w.r.t. all variables, which simplifies the subsequent optimization. 

{\textit{\textbf{Remark 2:} The reformulation approximation may incur error from two sources: (i) the rank-one recovery of $\widetilde{\mathbf{R}}_s$ via decomposition, which is mainly from the potential suboptimality of Gaussian randomization, and (ii) the imperfect realization of $\mathbf{s}_r$ via $\boldsymbol{\Theta}^\mathrm{H} \mathbf{G} \mathbf{x}$ under the joint ISAC design. The dominant error arises from the mismatch $\|\mathbf{s}_r - \boldsymbol{\Theta}^\mathrm{H} \mathbf{G} \mathbf{x}\|^2$, which is minimized. Although theoretical bounds are difficult to obtain, the approximation error remains trivial in practice.}}

Ultimately, (\ref{formulation}) is reformulated as
\begin{subequations}\label{formulation1}
\begin{align} 
\min _{\mathbf{x}, \omega,  \mathbf{p}, \boldsymbol{\Theta}} &~  \varepsilon=\alpha \left(\|\mathbf{s}_r-\boldsymbol{\Theta}^\mathrm{H} \mathbf{G} \mathbf{x}\|^2\right)\notag \\
& +  (1\!- \!\alpha) \left(\|\mathbf{s}_c \!-\!  \omega\mathbf{H}_{rc}^{\mathrm{H}} \boldsymbol{\Theta}^\mathrm{H} \mathbf{G} \mathbf{x}\|^2 \!+\! K \omega^2 \sigma_0^2\right) \tag{\ref{formulation1}{a}} \\ 
\mathrm{s.t.} &~  \mathrm{(\ref{formulation}{b})- (\ref{formulation}{f})}. \tag{\ref{formulation1}{b}}
\end{align}
\end{subequations}
 Subsequently, we solve  (\ref{formulation1}) by iteratively updating the variables. %It is worth mentioning that $\mathbf{s}_r$ may change after the update of $\mathbf{p}$, but according to the Lyapunov stability \cite{stability}, $\mathbf{s}_r$ is asymptotically stable over iterations due to the bounded $\mathbf{p}$ and the finite resources. 

\subsection{Optimization of Communication Symbol Estimator $\omega$} \label{sec:optimize omega}
Concerning $\omega$,  (\ref{formulation1}{a}) is a quadratic function. After ignoring the terms that are unrelated to $\omega$, (\ref{formulation1}) is simplified as
\begin{align} \label{optimize_omega}
    \min _{\omega \in \mathbb{R}} &~ ~ \varepsilon_c=\|\mathbf{s}_c -  \omega\mathbf{H}_{rc}^{\mathrm{H}} \boldsymbol{\Theta}^\mathrm{H} \mathbf{G} \mathbf{x}\|^2 + K \omega^2 \sigma_0^2.
\end{align}
The optimal $\omega^\star$ can be derived by setting the derivative of $\varepsilon_c$ to be zero, that is, $\frac{\partial \varepsilon_c}{\partial \omega}=0$, yielding the closed-form solution:
\begin{align} \label{opt_omega}
     \omega^\star  = \frac{\mathfrak{R}\lbrace\mathbf{s}^{\mathrm{H}}_c \mathbf{H}_{rc}^{\mathrm{H}} \boldsymbol{\Theta}^\mathrm{H} \mathbf{G} \mathbf{x}\rbrace}{\mathbf{x}^{\mathrm{H}} \mathbf{G}^{\mathrm{H}} \boldsymbol{\Theta} \mathbf{H}_{rc} \mathbf{H}_{rc}^{\mathrm{H}} \boldsymbol{\Theta}^\mathrm{H} \mathbf{G} \mathbf{x}+ K  \sigma_0^2}.
\end{align}

\subsection{Optimization of fRIS Phase Shift $\boldsymbol{\Theta}$} \label{sec:optimize FRIS}
Given $\mathbf{x}$, $\mathbf{p}$, and $\omega$,  (\ref{formulation1}) can be reduced to a modulus-one constrained quadratic minimization problem,  given by
\begin{subequations} \label{optimize_theta}
 \begin{align}
    \min _{ \boldsymbol{\Theta}} &~  \alpha \|\mathbf{s}_r \!-\! \boldsymbol{\Theta}^\mathrm{H} \mathbf{G} \mathbf{x}\|^2 \!+\!(1\!- \!\alpha) \|\mathbf{s}_c -  \omega\mathbf{H}_{rc}^{\mathrm{H}} \boldsymbol{\Theta}^\mathrm{H} \mathbf{G} \mathbf{x}\|^2. \\ 
\mathrm{s.t.} &~   |\boldsymbol{\Theta}|_{n,n}=1,  ~~\forall ~ n=1,\cdots, N.
\end{align}   
\end{subequations}
 To address (\ref{optimize_theta}), we first define $\boldsymbol{\theta}=\mathrm{diag}(\boldsymbol{\Theta})$ and employ the property of $\mathrm{tr}(\boldsymbol{\Theta}^\mathrm{H}\mathbf{C}_1\boldsymbol{\Theta}\mathbf{C}_2)=\boldsymbol{\theta}^\mathrm{H}(\mathbf{C}_1\odot\mathbf{C}_2^\mathrm{T})\boldsymbol{\theta}$ \cite{YJJ_EE} to transform 
(\ref{optimize_theta}) to a compact form:
\begin{subequations} \label{optimize_theta2}
 \begin{align}
    \min _{ \boldsymbol{\theta}} &~   \boldsymbol{\theta}^\mathrm{H} {\mathbf{A}}_1\boldsymbol{\theta}  - \boldsymbol{\theta}^\mathrm{H} {\mathbf{b}}_1 -  {\mathbf{b}}_1^\mathrm{H}  \boldsymbol{\theta} \\
\mathrm{s.t.} &~   |\boldsymbol{\theta}|_{n}=1, \forall n=1,\cdots, N.
\end{align}   
\end{subequations}
where  ${\mathbf{b}}_1 \!=\! \alpha \mathrm{diag}(\mathbf{G} \mathbf{x}  \mathbf{s}_r^\mathrm{H}) \!+ \omega (1\!- \!\alpha)  \mathrm{diag}(\mathbf{G} \mathbf{x}\mathbf{s}_c^{\mathrm{H}} \mathbf{H}_{rc}^{\mathrm{H}} )$ and ${\mathbf{A}}_1 \!= \!\alpha\left[\left(\mathbf{G} \mathbf{x} \mathbf{x}^\mathrm{H} \mathbf{G}^\mathrm{H} \right) \odot \mathbf{I}\right] + \omega^2  \left(1 \!-\!  \alpha\right) \! \left[\left(\mathbf{G} \mathbf{x} \mathbf{x}^\mathrm{H} \mathbf{G}^\mathrm{H} \right)  \! \odot \! \left(\mathbf{H}_{rc}^* \mathbf{H}_{rc}^{\mathrm{T}} \right) \right]$.
% \begin{subequations}
%   \begin{align}
%     {\mathbf{A}}_1 & \!= \!\alpha\left[\left(\mathbf{G} \mathbf{x} \mathbf{x}^\mathrm{H} \mathbf{G}^\mathrm{H} \right) \odot \mathbf{I}\right] \notag
%     \\
%     &~~ + \omega^2  \left(1 \!-\!  \alpha\right) \! \left[\left(\mathbf{G} \mathbf{x} \mathbf{x}^\mathrm{H} \mathbf{G}^\mathrm{H} \right)  \! \odot \! \left(\mathbf{H}_{rc}^* \mathbf{H}_{rc}^{\mathrm{T}} \right) \right], \\
%     {\mathbf{b}}_1& = \alpha \mathrm{diag}(\mathbf{G} \mathbf{x}  \mathbf{s}_r^\mathrm{H}) + \omega (1- \alpha)  \mathrm{diag}(\mathbf{G} \mathbf{x}\mathbf{s}_c^{\mathrm{H}} \mathbf{H}_{rc}^{\mathrm{H}} ).
% \end{align}  
% \end{subequations}
To address (\ref{optimize_theta2}{b}),  various approaches can be adopted, such as SDR \cite{SDR_RIS} and manifold optimization \cite{manifold}.  {While SDR produces suboptimal solutions due to rank-one relaxation, it ensures a worst-case $\pi/4$-approximation. Manifold optimization performs gradient descent over the complex circle manifold, transforming the problem into an unconstrained search. Despite inherent non-convexity, both methods achieve high-quality local optima and demonstrate strong empirical performance.}

\subsection{Optimization of Transmit Waveform $\mathbf{x}$} \label{sec:optimize_TBF}
In this subsection, the optimization of the transmit waveform $\mathbf{x}$ is discussed. First, (\ref{formulation1}) w.r.t. $\mathbf{x}$  is simplified as
\begin{subequations} \label{optimize_W}
\begin{align} 
\min _{\mathbf{x}} &~  \alpha \! \left(\! \|\mathbf{s}_r\! -\! \boldsymbol{\Theta}^\mathrm{H} \mathbf{G} \mathbf{x} \|^2\right) \! \! +\!  (1\!- \!\alpha) \! \left(\|\mathbf{s}_c \!-\!  \omega \mathbf{H}_{c} \mathbf{x} \|^2 \right) \\ 
\mathrm{s.t.} &~  \|\mathbf{x}\|^2 \leq P_t,
\end{align}
\end{subequations}
where $\mathbf{H}_{c}=\mathbf{H}_{rc}^{\mathrm{H}} \boldsymbol{\Theta}^\mathrm{H} \mathbf{G} \in \mathbb{C}^{K \times M}$ is denoted as the overall channel from the transmitter to the users. {For (\ref{optimize_W}{a}), the squared norm is expanded, and the constant terms are omitted, leading to the equivalent expression as
\begin{align} \label{objective_function_W}
    & \alpha \! \left( \mathbf{x}^\mathrm{H} \mathbf{G}^\mathrm{H} 
 \boldsymbol{\Theta}  \boldsymbol{\Theta}^\mathrm{H} \mathbf{G}\mathbf{x} - 2 \mathfrak{R}\lbrace\mathbf{s}_r^\mathrm{H} \boldsymbol{\Theta}^\mathrm{H} \mathbf{G} \mathbf{x}\rbrace \right) \notag \\
     &  \!\!\! + (1\!- \!\alpha) \! \left(\omega^2  \mathbf{x}^\mathrm{H} \mathbf{H}_{c}^\mathrm{H}   \mathbf{H}_{c} \mathbf{x} - 2 \omega \mathfrak{R}\lbrace\mathbf{s}_c^\mathrm{H} \mathbf{H}_{c} \mathbf{x} \rbrace \right). 
\end{align}
The objective function of (\ref{objective_function_W}) is compactly rewritten as 
\begin{align} \label{simplified_obj_w}
    \mathbf{x}^\mathrm{H} \mathbf{A}_2 \mathbf{x} - 2  \mathfrak{R}\lbrace \mathbf{b}_2^\mathrm{H}\mathbf{x} \rbrace
\end{align}
by denoting
\begin{subequations}\label{denotation_W}
\begin{align} 
    \mathbf{A}_2 & \!=\! \alpha \!\left( \!\mathbf{G}^\mathrm{H}\boldsymbol{\Theta}  \boldsymbol{\Theta}^\mathrm{H} \mathbf{G}\!\right) \!+\! (1\!- \!\alpha) \omega^2\!\left(\!\mathbf{H}_{c}^\mathrm{H}   \mathbf{H}_{c}\right)\!, \\
    \mathbf{b}_2 &= \alpha \left(\mathbf{G}^\mathrm{H}\boldsymbol{\Theta} \mathbf{s}_r  \right) + (1\!- \!\alpha) \omega \left(\mathbf{H}_{c}^\mathrm{H}\mathbf{s}_c\right).
\end{align}  
\end{subequations}
Notably, when $\boldsymbol{\Theta}$ is an ideal modulus-one diagonal matrix,   $\mathbf{G}^\mathrm{H} \boldsymbol{\Theta} \boldsymbol{\Theta}^\mathrm{H} \mathbf{G}$ in (\ref{denotation_W}{b}) can be reduced to $\mathbf{G}^\mathrm{H} \mathbf{G}$ due to $\boldsymbol{\Theta} \boldsymbol{\Theta}^\mathrm{H} = \mathbf{I}$.} After this transformation, it is seen that (\ref{simplified_obj_w}) is a convex quadratic function. For (\ref{optimize_W}{b}), it can be re-expressed as the inner product of $\mathbf{x}$, that is, $\mathbf{x}^\mathrm{H}\mathbf{x}\leq P_t$. In this context, we consider full power usage, namely, $\mathbf{x}^\mathrm{H}\mathbf{x}= P_t$. Therefore, (\ref{optimize_W}) is rewritten as
\begin{align} \label{optimize_W2}
\min _{\mathbf{x}} &~  \mathbf{x}^\mathrm{H} \mathbf{A}_2 \mathbf{x} - 2  \mathfrak{R}\lbrace \mathbf{b}_2^\mathrm{H}\mathbf{x} \rbrace ~~~~\mathrm{s.t.}  ~  \mathbf{x}^\mathrm{H}\mathbf{x}= P_t,
\end{align}
which can be addressed by ALM \cite{lxp_optimization,boyd}. Specifically, we first employ the complex-real conversion by separating the real part and the imaginary part. Thus, (\ref{optimize_W2}) is converted to its real form as
\begin{align} \label{optimize_W2_real}
\min _{\overline{\mathbf{x}}} &~  \mathring{f}(\overline{\mathbf{x}})=\overline{\mathbf{x}}^\mathrm{T} \overline{\mathbf{A}}_2 \overline{\mathbf{x}} - 2   \overline{\mathbf{b}}_2^\mathrm{T}\overline{\mathbf{x}} ~~~~
\mathrm{s.t.} ~  \overline{\mathbf{x}}^\mathrm{T}\overline{\mathbf{x}}= P_t,
\end{align}
where $ \overline{\mathbf{x}}=\left[ \mathfrak{R}\lbrace\mathbf{x^\mathrm{T}}\rbrace,\mathfrak{I}\lbrace\mathbf{x^\mathrm{T}}\rbrace\right]^\mathrm{T}$, $\overline{\mathbf{b}}_2=\left[\mathfrak{R}\lbrace\mathbf{x^\mathrm{T}}\rbrace ,\mathfrak{I}\lbrace\mathbf{x^\mathrm{T}}\rbrace\right]^\mathrm{T}$ and $\overline{\mathbf{A}}_2 =\left[\mathfrak{R}\lbrace\mathbf{A}_2\rbrace, -\mathfrak{I}\lbrace\mathbf{A}_2\rbrace;\mathfrak{I}\lbrace\mathbf{A}_2\rbrace, \mathfrak{R}\lbrace\mathbf{A}_2\rbrace\right]$.
% \begin{subequations}
% \begin{align}
%     \overline{\mathbf{w}}&=\begin{bmatrix}    \mathfrak{R}\lbrace\mathbf{w}\rbrace \\ \mathfrak{I}\lbrace\mathbf{w}\rbrace
% \end{bmatrix}, 
% \overline{\mathbf{b}}_2=\begin{bmatrix}    \mathfrak{R}\lbrace\mathbf{b}_2\rbrace \\ \mathfrak{I}\lbrace\mathbf{b}_2\rbrace
% \end{bmatrix},\\
% \overline{\mathbf{A}}_2&=\begin{bmatrix}    \mathfrak{R}\lbrace\mathbf{A}_2\rbrace & -\mathfrak{I}\lbrace\mathbf{A}_2\rbrace \\
% \mathfrak{I}\lbrace\mathbf{A}_2\rbrace& \mathfrak{R}\lbrace\mathbf{A}_2\rbrace
% \end{bmatrix}.
% \end{align}    
% \end{subequations}
Accordingly, the augmented Lagrangian function of  (\ref{optimize_W2_real}) can be written as
\begin{align} \label{lag}
    \mathcal{L} \triangleq \mathring{f}(\overline{\mathbf{x}})+ \mu \left(\overline{\mathbf{x}}^\mathrm{T}\overline{\mathbf{x}}- P_t\right) + \frac{\gamma}{2}\left(\overline{\mathbf{x}}^\mathrm{T}\overline{\mathbf{x}}- P_t\right)^2,
\end{align}
in which $\mu $ is the Lagrange multiplier and $\gamma$ denotes the penalty factor. Then, we tackle (\ref{lag}) using the following iteration procedure:
\begin{subequations} \label{update_w}
\begin{align}
    \overline{\mathbf{x}}^{(t)} &= \arg \min_{\overline{\mathbf{x}}}  \mathcal{L}(\overline{\mathbf{x}},\mu^{(t-1)}), \\
    \mu^{(t)} &= \mu^{(t-1)} + \gamma\left(\overline{\mathbf{x}}^{\mathrm{T }(t)}\overline{\mathbf{x}}^{(t)}- P_t\right).
\end{align}    
\end{subequations}
Specifically, we can solve (\ref{update_w}{a}) via the quasi-Newton method \cite{boyd}, which is implemented by the \textit{fminunc} tool. This process is terminated when  $\left|\overline{\mathbf{x}}^{\mathrm{T }}\overline{\mathbf{x}}- P_t\right|<\epsilon_0$ is satisfied. 

\subsection{Optimization of fRIS Element Positions $\mathbf{p}$} \label{sec:optimize_FRIS_pos}
In this subsection, we focus on the optimization of the fRIS element position $\mathbf{p}$. First,  (\ref{formulation1}) is simplified by omitting the unrelated terms and constraints:
\begin{subequations} \label{optimize_p}
\begin{align}
    \min _{\mathbf{p}} &~  \alpha \|\mathbf{s}_r-\boldsymbol{\Theta}^\mathrm{H} \mathbf{G}(\mathbf{p}) \mathbf{x}\|^2  \notag \\
    &~ + (1- \alpha) \|\mathbf{s}_c -  \omega\mathbf{H}_{rc}^{\mathrm{H}}(\mathbf{p}) \boldsymbol{\Theta}^\mathrm{H} \mathbf{G}(\mathbf{p}) \mathbf{x}\|^2  \\ 
    \mathrm{s.t.} &~   (\ref{formulation}{e})-(\ref{formulation}{f}) .
\end{align}  
\end{subequations}
Inspecting (\ref{optimize_p}), $\mathbf{G}$ and $\mathbf{H}_{rc}$ are determined by the variable $\mathbf{p}$, and (\ref{optimize_p}{a}) is a complicated composite function, where quadratic and quartic forms exist simultaneously in the outer function. In the interior function, $\mathbf{p}$ is contained in the complex exponential function, shown in (\ref{FA_steering_vector}). 

To explicitly observe $\mathbf{p}$ in (\ref{optimize_p}{a}), we  substitute $\mathbf{G}$ in (\ref{chan_G}) and $\mathbf{H}_{rc}$ in (\ref{chan_hrc}) to (\ref{optimize_p}{a}) and integrate the unrelated symbols of $\mathbf{p}$, which is given by the following proposition.

 \textbf{Proposition 1.} \textit{After omitting the constants, (\ref{optimize_p}{a}) is  rewritten as
        \begin{align} \label{p_obj1}
            \widetilde{f}_0 =& -2 \alpha \sqrt{\zeta_G} \mathfrak{R}\left\lbrace \widetilde{\mathbf{a}}_t^\mathrm{H}  \mathbf{a}_r(\mathbf{p}) \right\rbrace \notag \\
             &  +(1\!-\! \alpha)\left[ c_0 \sum_{k=1}^{K}  \zeta_k   \mathbf{a}_r^{\mathrm{H}}(\mathbf{p})   \widetilde{\mathbf{A}}_{c,k}( \mathbf{p}) \boldsymbol{\theta}    \boldsymbol{\theta}^\mathrm{H} \widetilde{\mathbf{A}}_{c,k}^\mathrm{H}(\mathbf{p}) \mathbf{a}_r(\mathbf{p})  \right. \notag \\
            & \left. - 2 \omega \sqrt{\zeta_G} \mathfrak{R}\left \lbrace  \mathrm{tr}\left(\mathbf{S} \mathbf{A}_{rc}^{\mathrm{H}} (\mathbf{p}) 
  \widetilde{\mathbf{A}}_{r}( \mathbf{p})  \right)\right\rbrace \right],
        \end{align}
   where we denote $\widetilde{\mathbf{a}}_t^\mathrm{H} = \mathbf{a}_t ^\mathrm{H} \mathbf{x}\mathbf{s}_r^\mathrm{H}\boldsymbol{\Theta}^\mathrm{H}$, $c_0 = \omega^2 \zeta_G c_x$, $c_x = \mathbf{a}_t ^\mathrm{H}\mathbf{x}\mathbf{x}^{\mathrm{H}}\mathbf{a}_t$, $\widetilde{\mathbf{A}}_{c,k}( \mathbf{p}) = \mathrm{Diag}(\mathbf{a}_{c,k}( \mathbf{p}))$, $\widetilde{\mathbf{A}}_{r}( \mathbf{p}) = \mathrm{Diag}(\mathbf{a}_{r}( \mathbf{p}))$ and $\mathbf{S}=\boldsymbol{\theta}^* \mathbf{a}_t^{\mathrm{H}} \mathbf{x} \mathbf{s}_c^{\mathrm{H}}  \boldsymbol{\Sigma}_{rc}^{\mathrm{H}}$.}

\textit{Proof:} The proof is provided in Appendix \ref{app_a}.

However, $\mathbf{p}$ in (\ref{p_obj1}) is contained in the array manifold, while there exist $K$ quartic terms, a quadratic term, and a linear item w.r.t. the array manifold. Fortunately, it is shown in the following proposition that these items can be reduced to the first-order items and the difference of the first-order items if we focus on optimizing the position of one element at a time. 

 {\textbf{Proposition 2.} }  \textit{When considering the position of the $n$-th element $\mathbf{p}_n$ while fixing others, $\widetilde{f}_0$ in (\ref{p_obj1}) is simplified as
   \begin{align} \label{p_obj2}
    \widetilde{f}_1 (\mathbf{p}_n) &= \widetilde{f}_{1,1}+\widetilde{f}_{1,2}+\widetilde{f}_{1,3} \notag \\
    &= -\nu_{n}  \cos \left( \xi_n +\frac{2\pi}{\lambda}d_{r,n}(\mathbf{p}_n) \right) \notag \\   
    & ~~+ \sum_{k=1}^{K} \sum_{i=1, i\neq n}^{N} \widetilde{\nu} \zeta_{k}  \cos \left( \widetilde{\xi}_{i,n,k} +  \frac{2\pi}{\lambda} \Delta \widetilde{d}_{k,n} (\mathbf{p}_n) \right) \notag \\
    &~~ - \sum_{k=1}^{K}  \overline{\nu} ~\overline{\rho}_{n,k} \cos\!\left (  \overline{\xi}_{n,k} \!+\! \frac{2\pi}{\lambda}\Delta \widetilde{d}_{k,n}(\mathbf{p}_n)  \right).
\end{align}
In the first term, we denote $\xi_n$ and $\rho_n$ as the phase and amplitude of the $n$-th entry of $~\widetilde{\mathbf{a}}_t^\mathrm{H}$, while  $\nu_{n}\!=\! 2\alpha  \sqrt{\zeta_G} \rho_n$ is the constant. The symbol of $d_{r,n}(\mathbf{p}_n)$ is defined as $p_{x,n} \sin (\phi_r)\cos (\psi_r) \!+\! p_{y,n} \sin (\psi_r)$. In the second term, $\Delta \widetilde{d}_{k,n}(\mathbf{p}_n) \!=\!d_{r,n}(\mathbf{p}_n)\!-\!d_{c,k,n}(\mathbf{p}_n)$, while $\widetilde{\xi}_{i,n,k}$ is the phase of a constant $\widetilde{c}_{i,n,k}$. Meanwhile, we denote $\widetilde{\nu}=2(1\!-\! \alpha) c_0$. In the last term, we represent $\overline{\xi}_{n,k}$ as the phase of $s_{n,k}$, while $\overline{\nu}=2 (1\!-\! \alpha) \omega \sqrt{\zeta_G}$.}

\textit{Proof:} The proof is provided in Appendix \ref{app_b}.

{\textit{\textbf{Remark 3:} The derivations in Propositions 1 and 2 rely on structural assumptions of both the fRIS and the channel model. Specifically, the fRIS is required to be passive and single-connected, i.e., \( |\boldsymbol{\Theta}_{i,i}| = 1 \). Furthermore, the channels are considered as LoS channels, which is reasonable in practical scenarios.  The RIS is strategically deployed to establish virtual LoS links, for example, to bypass obstacles or extend coverage.}}

Observing $\widetilde{f}_1$ in (\ref{p_obj2}), it contains cosine functions, which are neither convex nor concave w.r.t. $\mathbf{p}_n$. To address this problem, we employ the MM framework, where surrogate functions that upper bound $\widetilde{f}_1$ are constructed. Specifically, the construction of the surrogate functions is given by the following lemma. 

 {\textbf{Lemma 1. \cite{lemma_proof}} } \textit{Considering a function $\widetilde{f}$ with $\mathbf{z}$ being the variable, its second-order Taylor expansion at $\mathbf{z}_0$ is given by
\begin{align}
      \widetilde{f}(\mathbf{z})\approx   \widetilde{f}(\mathbf{z}_0) &+ (\nabla \widetilde{f}(\mathbf{z}_0))^\mathrm{T} (\mathbf{z}-\mathbf{z}_0) \notag\\
      &+ \frac{1}{2} (\mathbf{z}-\mathbf{z}_0)^\mathrm{T} \nabla^2\widetilde{f}(\mathbf{z}_0)  (\mathbf{z}-\mathbf{z}_0),
\end{align}
where $\nabla \widetilde{f}(\mathbf{z})$ is the gradient vector of $\widetilde{f}(\mathbf{z})$, while $\nabla^2\widetilde{f}(\mathbf{z})$ represents the Hessian matrix of $\widetilde{f}(\mathbf{z})$. }

\textit{By selecting a parameter $\delta$ satisfying $\delta \mathbf{I} \succeq \nabla^2 \widetilde{f}(\mathbf{z}_0)$, the function $ \widetilde{f}(\mathbf{z}) $ can be upper-bounded by
\begin{align}
    \!\widetilde{f}(\mathbf{z})  \leq  \widetilde{f}(\mathbf{z}_0) \! +\! (\nabla \widetilde{f}(\mathbf{z}_0))^\mathrm{T} (\mathbf{z} - \mathbf{z}_0)  \!+ \! \frac{\delta}{2} \left\| (\mathbf{z} - \mathbf{z}_0)\right \|^2. 
\end{align}
In particular,  $\delta$ is chosen as the Frobenius norm of   $\nabla^2\widetilde{f}(\mathbf{z}_0)$.}

According to  \textbf{Lemma 1}, we separate $\mathbf{p}_n$ from the cosine functions and utilize the second-order Taylor expansion as the surrogate function. To construct the surrogate function, the gradient vector and the Hessian matrix of  $\widetilde{f}_1(\mathbf{p}_n)$ should be determined, which are given by the following theorem.

 {\textbf{Theorem 1.} } \textit{According to the linear property of derivative operation, the gradient of $\widetilde{f}_1(\mathbf{p}_n)$ is the sum of the derivatives of its three components, which is expressed as
\begin{align} \label{Gradient}
    \nabla \widetilde{f}_1 (\mathbf{p}_n) = \begin{bmatrix}
        \sum_{i=1}^{3}\frac{\partial \widetilde{f}_{1,i}(\mathbf{p}_n)}{\partial p_{x,n}}  &  \sum_{i=1}^{3} \frac{\partial \widetilde{f}_{1,i}(\mathbf{p}_n)}{\partial p_{y,n}}
    \end{bmatrix}^\mathrm{T}.
\end{align}
For the Hessian matrix of $\widetilde{f}_1$, it is given by
\begin{align} \label{Hessian}
    \nabla^2 \widetilde{f}_1 (\mathbf{p}_n) = \begin{bmatrix}
        \sum_{i=1}^{3}\frac{\partial^2 \widetilde{f}_{1,i}(\mathbf{p}_n)}{\partial p^2_{x,n}}  &  \sum_{i=1}^{3} \frac{\partial^2 \widetilde{f}_{1,i}(\mathbf{p}_n)}{\partial p_{x,n}\partial p_{y,n}}\\
         \sum_{i=1}^{3}\frac{\partial^2 \widetilde{f}_{1,i}(\mathbf{p}_n)}{\partial p_{y,n}\partial p_{x,n} } &  \sum_{i=1}^{3} \frac{\partial^2\widetilde{f}_{1,i}(\mathbf{p}_n)}{\partial p^2_{y,n}}
    \end{bmatrix}.
\end{align}
Specifically, there are 6 first-order derivative terms in (\ref{Gradient}) and 12 second-order derivative terms in (\ref{Hessian}), whose detailed expressions are left in the Appendix \ref{app_c} for brevity.}

\textit{Proof:} Please refer to Appendix \ref{app_c}.

According to \textbf{Lemma 1} and \textbf{Theorem 1}, the relaxed objective function is given by
\begin{align} \label{obj_p_final}
     (\nabla \widetilde{f}_1(\mathbf{p}_{n}^{(t-1)}))^\mathrm{T} (\mathbf{p}_n - \mathbf{p}_{n}^{(t-1)})  \!+ \! \frac{\delta_n}{2} \left\| (\mathbf{p}_n - \mathbf{p}_{n}^{(t-1)})\right \|^2,
\end{align}
where the constants are neglected, while $\delta_n=\|\nabla^2 \widetilde{f}_1 (\mathbf{p}_n)\|^2_\mathrm{F}$. The $\mathbf{p}_{n}^{(t-1)}$ represents the  position of the $n$-th element from the previous optimization result.  Apparently, (\ref{obj_p_final}) is a convex quadratic function w.r.t. $\mathbf{p}_n$. 

{ \textit{{\textbf{Remark 4}:} The construction of (\ref{obj_p_final}) guarantees that the surrogate and the original function match in both value and gradient at the current point $\mathbf{p}_n^{(t-1)}$, ensuring local approximation. In particular, the approximation error can be characterized as}
\begin{align}
    \text{error} 
    &\approx \frac{1}{2} \Delta\mathbf{p}_n^\mathrm{T} \left(\delta_n \mathbf{I}-\nabla^2\widetilde{f}_1(\mathbf{p}_{n}^{(t-1)}) \right )\Delta\mathbf{p}_n,
\end{align}
\textit{where $\Delta\mathbf{p}_n = \mathbf{p}_n - \mathbf{p}_n^{(t-1)}$. The upper-bounding strategy ensures that the surrogate becomes increasingly tight during convergence. Nevertheless, since the original problem is non-convex, the MM algorithm is only guaranteed to converge to a stationary point, which is generally a local optimum.}
}

Apart from the objective function,  (\ref{formulation}{f}) remains non-convex. To deal with it, we relax (\ref{formulation}{f}) with the first-order Taylor expansion. Specifically, the first-order Taylor expansion of the left term in (\ref{formulation}{f}) is given by
\begin{subequations}
\begin{align}
      \|\mathbf{p}_n - \mathbf{p}_{n'}\|_2 &\geq \|\mathbf{p}_n^{(t\!-\!1)} \!-\! \mathbf{p}_{n'}\|_2 +\! \frac{(\mathbf{p}_n^{(t\!-\!1)} \!-\! \mathbf{p}_{n'})^\mathrm{T}\! ( \mathbf{p}_{n}\!-\! \mathbf{p}_n^{(t\!-\!1)})}{\|\mathbf{p}_n^{(t-1)} \!-\! \mathbf{p}_{n'}\|_2} \\
     &=\frac{(\mathbf{p}_n^{(t\!-\!1)} \!-\! \mathbf{p}_{n'})^\mathrm{T}\! ( \mathbf{p}_{n}\!-\! \mathbf{p}_{n'})}{\|\mathbf{p}_n^{(t-1)} \!-\! \mathbf{p}_{n'}\|_2} 
\end{align}    
\end{subequations}
As a result, (\ref{formulation}{f}) is relaxed as
\begin{align} \label{constr_p}
    \!\frac{(\mathbf{p}_n^{(t\!-\!1)} \!-\! \mathbf{p}_{n'})^\mathrm{T}\! ( \mathbf{p}_{n}\!-\! \mathbf{p}_{n'})}{\|\mathbf{p}_n^{(t-1)} \!-\! \mathbf{p}_{n'}\|_2}  \!\geq \!\Delta D, \forall n,n'\!=\! 1:N, n\!\neq\! n' ,
\end{align}
which is affine regarding $\mathbf{p}_n$. {
Notably,  this method relaxes the non-convex feasible region to an inner half-space feasible region. Since part of the feasible region is discarded, the optimization may only find a suboptimal solution. Nonetheless, such approximation ensures algorithmic stability and is commonly accepted in MM-based optimizations.} Therefore, the problem for optimizing the position of the $n$-th element  is given by
\begin{subequations} \label{optimize_p_final}
\begin{align}
    &\!\!\!\!\min _{\mathbf{p}_n} ~  \nabla \widetilde{f}_1(\mathbf{p}_{n}^{(t-1)})^\mathrm{T} (\mathbf{p}_n \!-\! \mathbf{p}_{n}^{(t\!-\!1)}) + \! \frac{\delta_n}{2} \left\| (\mathbf{p}_n \!-\! \mathbf{p}_{n}^{(t\!-\!1)})\right \|^2  \\ 
    &\!\!\!\!\mathrm{s.t.} ~   (\ref{formulation}{e}), (\ref{constr_p}),
\end{align}    
\end{subequations}
which is a typical convex quadratic problem, and solved by using quadprog \cite{quadprog}. {Besides, we adopt the circle packing scheme \cite{initial} for the initialization of fRIS element positions, where the initial positions are sufficiently separated. This scheme guarantees minimum spacing and provides sufficient mobility space for subsequent iterations to avoid collision.} The details of the proposed algorithm are summarized in \textbf{Algorithm 1}. 

{\textit{\textbf{Remark 5}: Element repositioning may introduce additional overhead as compared to the traditional RIS. On the one hand, element displacement consumes energy, which mainly stems from the step motor drive or the port switching.  On the other hand, element mobility adds computational complexity to position optimizations. Nonetheless, the extra spatial DoFs of fRIS allow the use of fewer elements, keeping the total computational burden reasonable.}}

\begin{algorithm}[!t] 
\caption{Optimization of fRIS element positions} 
    \begin{algorithmic}[1] 
        \REQUIRE ~  Channel parameters, transmit signal, fRIS phase shift matrix,  reference signal, and communication symbols;
        \STATE Initialize $\mathbf{p}^{(0)}$ with the circle packing scheme;
        \FOR {$n=1\longrightarrow N$}
        \STATE Calculate the gradient and the Hessian matrix of $\widetilde{f}_1$ according to the \textbf{Theorem 1};
        \STATE Calculate $\delta_n$ based on the Hessian matrix of $\widetilde{f}_1$;
        \STATE Solve the problem (\ref{optimize_p_final}) using quadprog;
        \ENDFOR	
        \RETURN fRIS element positions $\mathbf{p}$. 
    \end{algorithmic}
\end{algorithm}

\subsection{Convergence and Computational Complexity Analysis} \label{conv_complex}
The proposed algorithm is summarized in \textbf{Algorithm 2}, which will terminate when the change of $\varepsilon$ falls below a threshold $\epsilon_0$ or the maximum iteration number $I_{\mathrm{max}}$ is reached.

\subsubsection{Convergence Analysis}
{The convergence behavior of the proposed \textbf{Algorithm 2} is analyzed as below. }

{\textbf{(i) Subproblem of symbol estimator optimization}: The update of $\omega$ in (\ref{optimize_omega}) solves a convex quadratic problem, which guarantees $\varepsilon(\omega^{t+1}, \mathbf{\Theta}^t, \mathbf{x}^t, \mathbf{p}^t) \leq \varepsilon(\omega^t, \mathbf{\Theta}^t, \mathbf{x}^t, \mathbf{p}^t)$. }

{\textbf{(ii) Subproblem of phase shift optimization}: The subproblem (\ref{optimize_theta2}) involves the modulus-one constraint. It is solved via either SDR \cite{SDR_RIS} or manifold optimization \cite{manifold}, where the SDR is emphatically shown as a non-increasing process and manifold optimization updates variables along the Riemannian gradient direction on the complex circle manifold. Both approaches ensure $\varepsilon(\omega^{t+1}, \mathbf{\Theta}^{t+1}, \mathbf{x}^t, \mathbf{p}^t) \leq \varepsilon(\omega^{t+1}, \mathbf{\Theta}^t, \mathbf{x}^t, \mathbf{p}^t)$.}

{\textbf{(iii) Subproblem of transmit waveform optimization}: In the optimization of $\mathbf{x}$, the quasi-Newton method guarantees that the update of $\overline{\mathbf{x}}$ is non-increasing \cite{boyd}, while the equality constraint is satisfied if the penalty parameter becomes sufficiently large. As a result,  the solution to (\ref{update_w}) can satisfy the equality constraint, while the process is non-increasing: $\varepsilon (\omega^{t+1}\!,  \mathbf{\Theta}^{t+1}\!, \mathbf{x}^{t+1}\!, \mathbf{p}^{t}) \!\leq\! \varepsilon (\omega^{t+1}\!, \mathbf{\Theta}^{t+1}\!, \mathbf{x}^{t}\!, \mathbf{p}^t)$.}

{\textbf{(iv) Subproblem of element position optimization}: The update of $\mathbf{p}$ is tackled via the MM framework, where a convex quadratic surrogate function is constructed. According to the convergence theory of MM methods \cite{MM}, the surrogate ensures that the original objective is non-increasing, i.e., 
    $\varepsilon(\mathbf{p}_n^{(i-1)})  = g(\mathbf{p}_n^{(i-1)}) +C \geq g(\mathbf{p}_n^{(i)}) +C =  \varepsilon(\mathbf{p}_n^{(i)})$, where $g(\cdot)$ is the surrogate function while $C$ is a constant. Besides, the inter-element spacing constraint is relaxed to a half space, which does not affect the monotonicity of the descent. }

{Moreover, due to the non-negativity of MSEs and the finite transmit power, the overall objective function is lower-bounded by $0$. Hence, the proposed \textbf{Algorithm 2} generates a non-increasing and bounded sequence of objective values,} {and  converges to a stationary point that is at least a local optimum.}

\subsubsection{Computational Complexity Analysis}
In this part, the computational complexity is analyzed. 
\begin{itemize}
    \item To obtain $\mathbf{s}_r$, the computational complexity mainly stems from solving (\ref{formulation0}) with SDR and the eigenvalue decomposition. Given a solution accuracy $\epsilon_1$, the computational complexity of SDR is $\mathcal{O}\left(N^{4.5} \log(1/\epsilon_1)\right)$ \cite{SDR}, while the eigenvalue decomposition has the computational complexity of $\mathcal{O}\left(N^3\right)$.
    \item  The optimal symbol estimator can be calculated via a closed-form expression, whose computational complexity stems from matrix multiplications. Obtaining the numerator in (\ref{opt_omega}) has the computation complexity of $\mathcal{O}\left(KN+N^2+MN+M\right)$, while calculating the denominator has the computation complexity of $\mathcal{O}\left(2(MK+NK+N^2)+M\right)$.
    \item When optimizing the fRIS phase shift, the computational complexity of the calculation of the matrix multiplication is $\mathcal{O}\left(MN+MK+N^2K+N^2M+MNK\right)$. Besides, the SDR approach has the computational complexity of  $\mathcal{O}\left(N^{4.5}\log(1/\epsilon_2)\right)$ with $\epsilon_2$ being the solution accuracy, while that of the manifold optimization is $\mathcal{O}\left(N^{1.5}\right)$ \cite{manifold}.
    \item In the optimization of transmit waveform design, the computational complexity of matrix multiplication is $\mathcal{O}\left(\!M^2N\!+\!M^2K\!+\!N^2M\!+\!MK\!+\!MN\!\right)$, while the design of the transmit waveform with ALM is $\mathcal{O}\left((M^2+M)\log\left(1/\epsilon_3\right)\right)$, where $\epsilon_3$ is the convergence accuracy \cite{ADMM_compl,quasinewton_compl}. % Also refer to (1) https://www.researchgate.net/post/what_is_the_computational_complexity_of_quasi-Newton_method_for_optimization
    % (2) https://scicomp.stackexchange.com/questions/42069/are-quasi-newton-methods-computationally-impractical
    % (3) https://infermatic.ai/ask/?question=What+are+the+computational+complexities+of+quasi-Newton+methods+in+high-dimensional+spaces%3F
    \item When optimizing the position of one fRIS element, the computational complexity of obtaining the gradients and the Hessian matrix is $\mathcal{O}\left(NK+K+M+N^2\right)$,  while the computational complexity of calculating $\delta_n$ is $\mathcal{O}\left(1\right)$. Additionally, the computational complexity of solving the problem (\ref{optimize_p_final}) is  $\mathcal{O}\left(\left(N^{1.5}\log(1/\epsilon_4)\right)\right)$ \cite{FA_Capacity} with accuracy $\epsilon_4$ for the interior-point
method. 
\end{itemize}
 In summary, the computational complexity of each iteration of the \textbf{Algorithm 2} can be approximated as $\mathcal{O}\left(N^{4.5} \log(1/\epsilon)\right)$, neglecting the lower-order terms. %\textcolor{red}{Therefore, the total algorithm complexity is approximately given by $\mathcal{O}\left( N^{4.5} \log(1/\epsilon)\cdot I_e\right)$, where $I_e$ is the number of iterations required for convergence. When the maximum number of iteration is reached, the computational complexity is given by $\mathcal{O}\left( N^{4.5} \log(1/\epsilon) \cdot I_{\mathrm{max}} \right)$.}%\mathcal{O}\left(N^{4.5} \log(1/\epsilon_1)+N^3+N^{1.5}+M^2\log\left(1/\epsilon_2\right)\right.$ $\left. +N\left( NK+N^{1.5}\log(1/\epsilon_3)\right)\right) \sim 

\begin{algorithm}[!t] 
\caption{Proposed alternating optimization algorithm} 
    \begin{algorithmic}[1] 
        \REQUIRE ~ Channel parameters, noise power, transmit power, ideal sensing beampattern, weighting coefficient,  fRIS size, and the minimum inter-element spacing.	
        \STATE Initialize all the variables, and $i=0$;
        \WHILE { {${|\varepsilon^{(i+1)}-\varepsilon^{(i)}|}/{\varepsilon^{(i)}}>\epsilon_0$ or  the maximum iteration number is not reached,}}
        \STATE Calculate correlation matrix by (\ref{formulation0}) and obtain reference signal $\mathbf{s}_r$ via (\ref{obtain_sr});
        \STATE Update communication symbol estimator by (\ref{opt_omega});
        \STATE Optimize $\boldsymbol{\Theta}$ by solving problem (\ref{optimize_theta2});
        \STATE Update ${\mathbf{x}}$ by implementing the ALM process in (\ref{update_w});
        \STATE Optimize fRIS element positions via Algorithm 1;
        \STATE {Update iteration index $i=i+1$};
        \ENDWHILE	
        \RETURN Optimized variables and joint MSE.
    \end{algorithmic}
\end{algorithm}

\section{Simulation Results} \label{simulation}
 This section provides numerical results to validate the effectiveness of the system and the proposed algorithm. We set the default parameters as follows.  The center frequency of this system is $2.4$GHz. The ISAC BS is located at ($3$, $0$, $0$)m, and equipped with  $M\!=\!8$ antennas. The transmitter has maximum transmit power of $10$dBm. The fRIS, consisting of  $25$ elements ($N_x\!=\!N_y=5$), is positioned at ($0$, $3$, $3$)m, with a size of $A \!\times\! A$ where  $A\!=\!5\lambda$ and a minimal inter-element spacing of  $\Delta D=\lambda/2$.  Additionally, we consider $K\!\!=\!\!4$ users and $T\!\!=\!\!3$ targets. The users are randomly distributed within an area centered at ($50$, $100$, $0$)m with a radius of $10$m. Observed from fRIS, the targets' (azimuth, elevation) angles are ($-20^\circ$, $0^\circ$), ($5^\circ$, $0^\circ$), and ($30^\circ$, $0^\circ$), respectively. The noise power is set as $\sigma_0^2\!=\!-60$dBm, and QPSK is used for modulation. The weighting coefficient  $\alpha$ is $0.5$. Moreover, the channel fading coefficients $\zeta_G$ and $\zeta_k$ are calculated by $\eta\cdot(\frac{1}{\text{dist}})^{2}$, where $\eta\!=\! -3$dB is the path loss, and dist denotes the transmitter-receiver distance \cite{SDR_RIS}. The penalty factor for ALM and the convergence threshold are set as $1$ and $10^{-5}$, respectively. All simulation results are averaged over $100$  Monte Carlo trials. {In addition,  $\mathbf{x}$, $\boldsymbol{\Theta}$, and $\omega$ are randomly initialized under their respective constraints, while the fRIS element positions $\mathbf{p}$ are initialized as stated in Section \ref{sec:optimize_FRIS_pos}.}

\begin{figure}[!t]
    \includegraphics[scale=0.53]{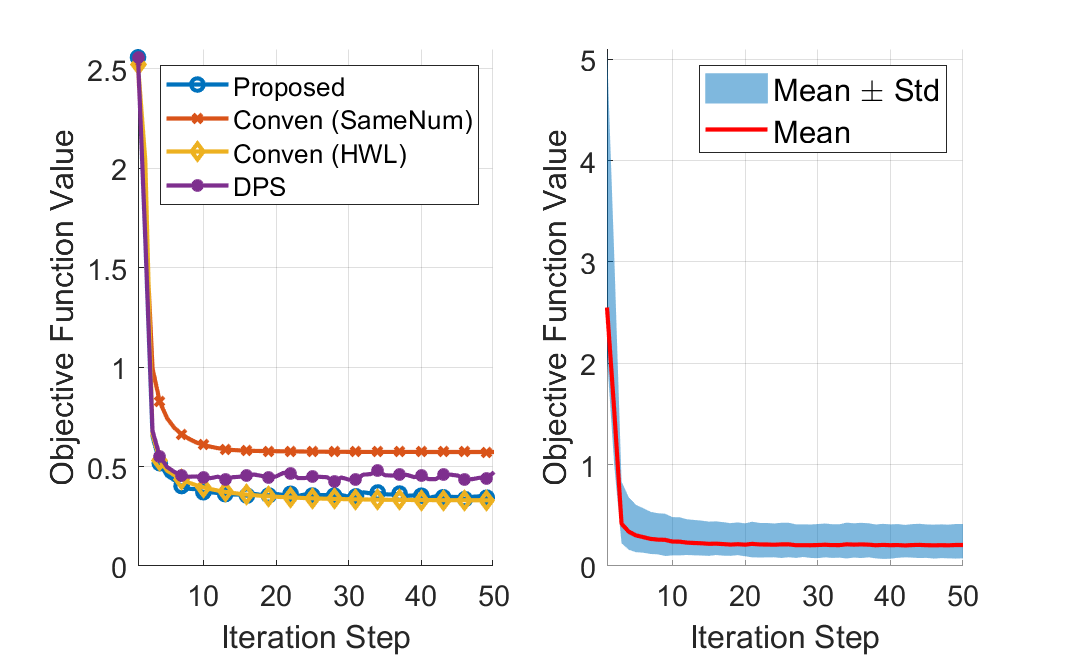}\\
    \centerline{\footnotesize{(a)}~~~~~~~~~~~~~~~~~~~~~~~~~~~~~~~~~~~~ (b)}  
    \caption{Convergence behaviors: (a) The proposed algorithm and the benchmarks: the objective function value $\varepsilon$ versus the iteration step. (b) Fluctuations with a fixed channel realization and random initialization parameters.}
    \label{fig:convergence}
\end{figure}

% \begin{figure}[!t]
%     \centering
%     \includegraphics[scale=0.4]{convergence.eps}
%     \caption{Convergence behaviors of the proposed algorithm and the benchmarks: the objective function value $\varepsilon$ versus the iteration step.}
%     \label{fig:convergence}
% \end{figure}

% \begin{figure}[!t]
%     \centering
%     \includegraphics[scale=0.4]{Initialization.eps}
%     \caption{Convergence fluctuations with a fixed channel realization and random initialization parameters.}
%     \label{fig:init}
% \end{figure}

The performance of the proposed design is evaluated against the following benchmark schemes:
\begin{itemize}
    \item \textbf{Conven (SameNum)}: This scheme is an ISAC aided by fixed-element RIS, where the RIS has the same element number as fRIS (sparse).  The rest of the beamforming designs follow the same procedures as proposed.
    
     \item \textbf{{Conven (HWL)}}: {The ISAC system is aided by a fixed-element RIS, where the elements have dense half-wavelength spacing in the same surface size.  The rest of the optimization procedures are identical to the proposed.}
     
     \item \textbf{DPS}: In this scheme, the fRIS element position is discrete, where the fRIS is quantized into discrete grids of equal spacing. Specifically, the continuous position is projected to the nearest position. 
 \end{itemize}

\begin{figure*}
	\centering
	\subfigure[]{\label{Fig_Beam_01}
		\includegraphics[height=3.6cm, width=4.3cm]{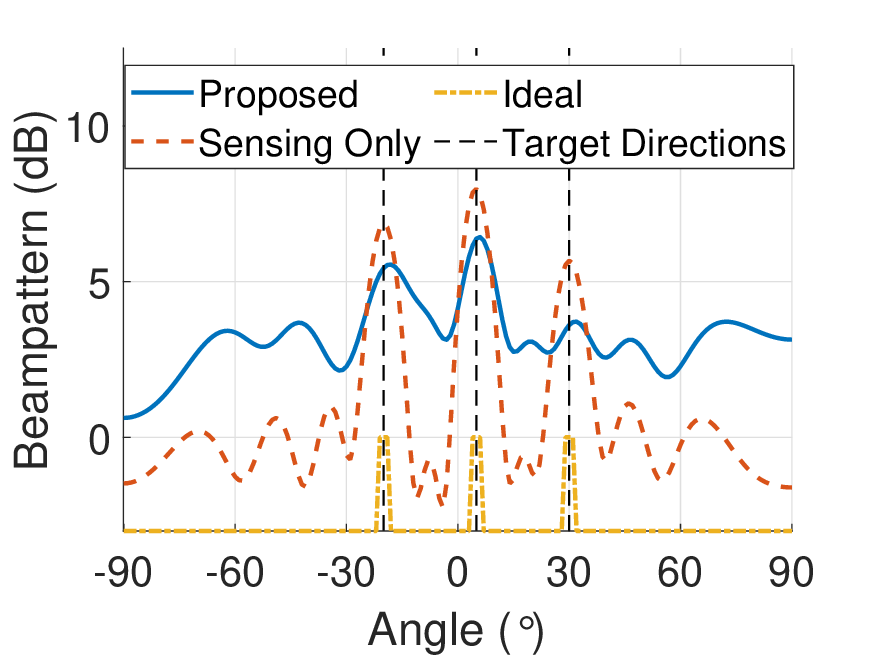}}
	\subfigure[]{\label{Fig_Beam_05}
		\includegraphics[height=3.6cm,width=4.3cm]{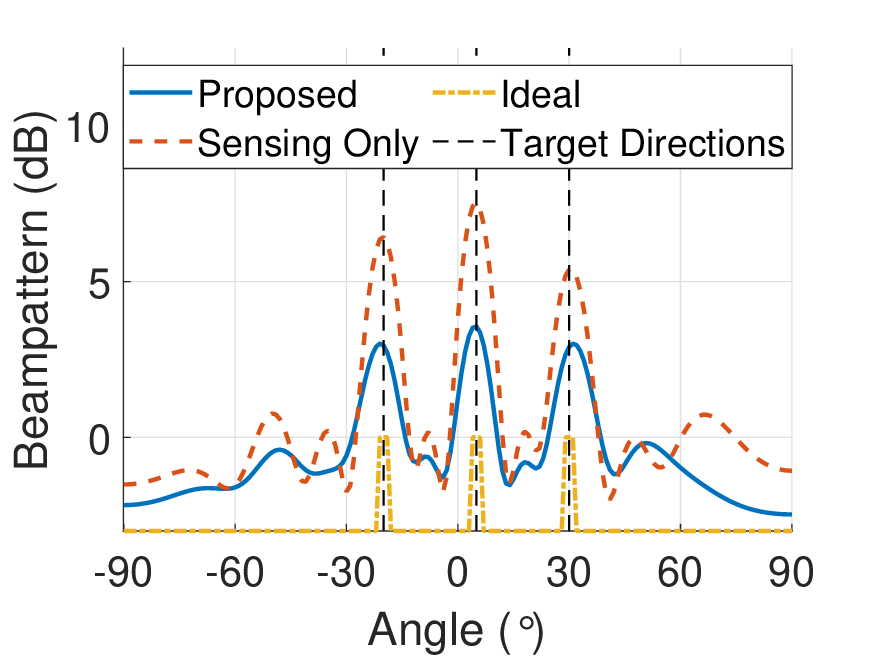}}	
	\centering
	\subfigure[]{\label{Fig_Beam_09}
		\includegraphics[height=3.6cm,width=4.3cm]{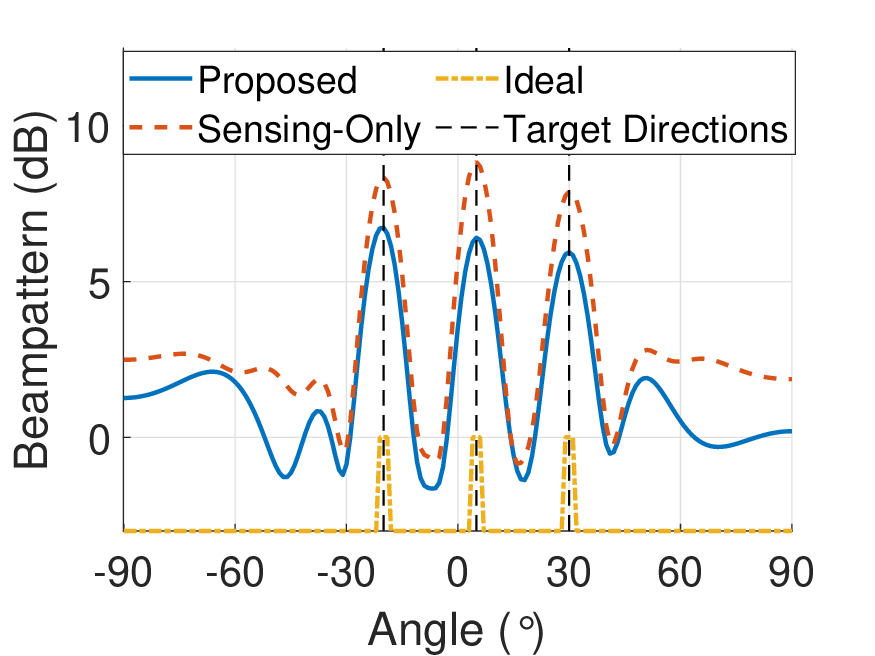}}
	\centering
	\subfigure[]{\label{Fig_Beam_Comp}
		\includegraphics[height=3.6cm,width=4.3cm]{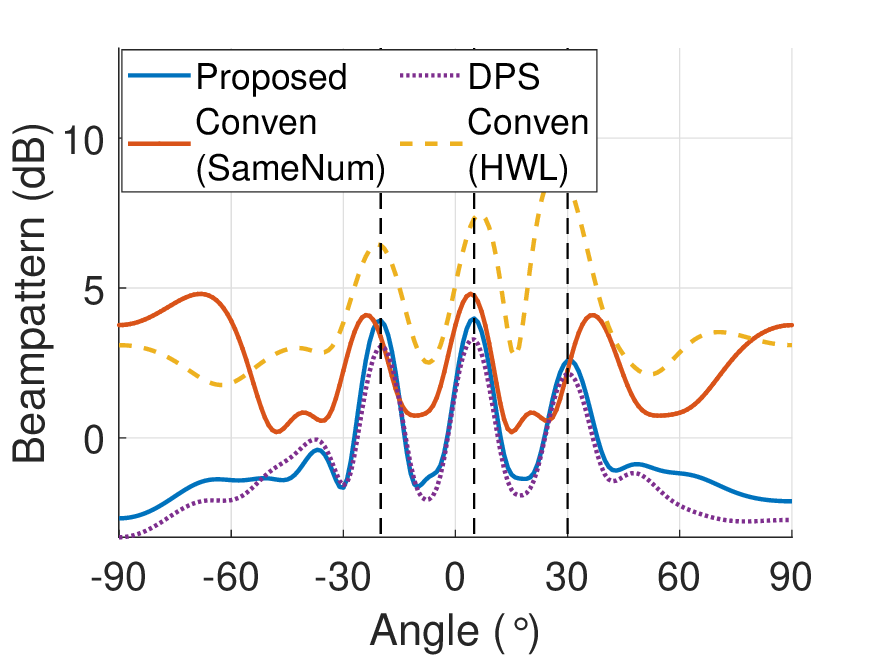}}	
	\caption{Beampattern at the fRIS: beam gain (dB) versus spatial angle ($^\circ$). (a) The beampattern when biased to the communication ($\alpha=0.1$). (b) The beampattern when no bias ($\alpha=0.5$). (c) The beampattern when biased to the sensing ($\alpha=0.9$). (d) The beampattern of different approaches when $\alpha=0.5$. }
	\label{Fig_Beams}		
\end{figure*}

\begin{figure*}
	\centering
	\subfigure[]{
		\includegraphics[width=5.2cm]{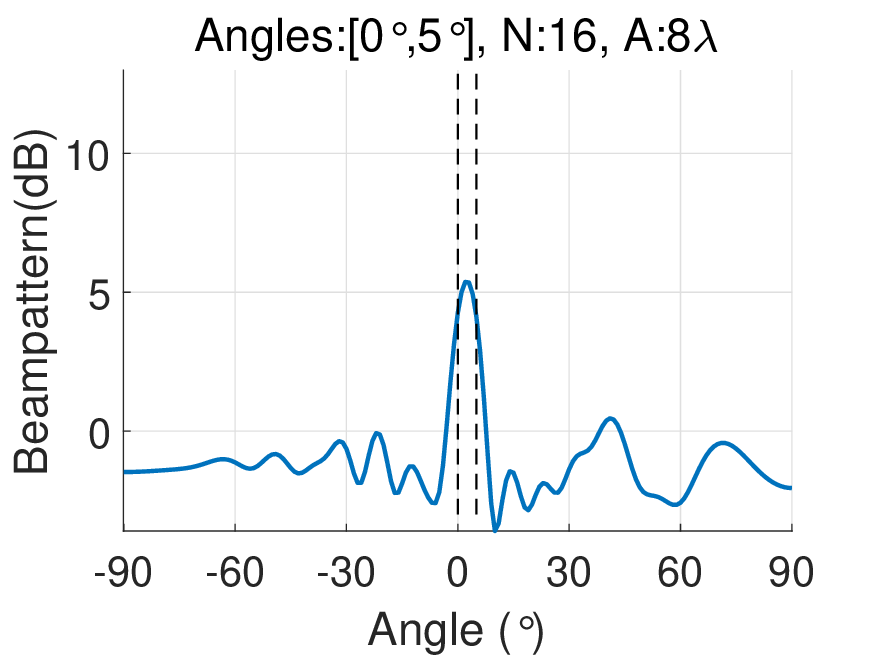}}
	\subfigure[]{
		\includegraphics[width=5.2cm]{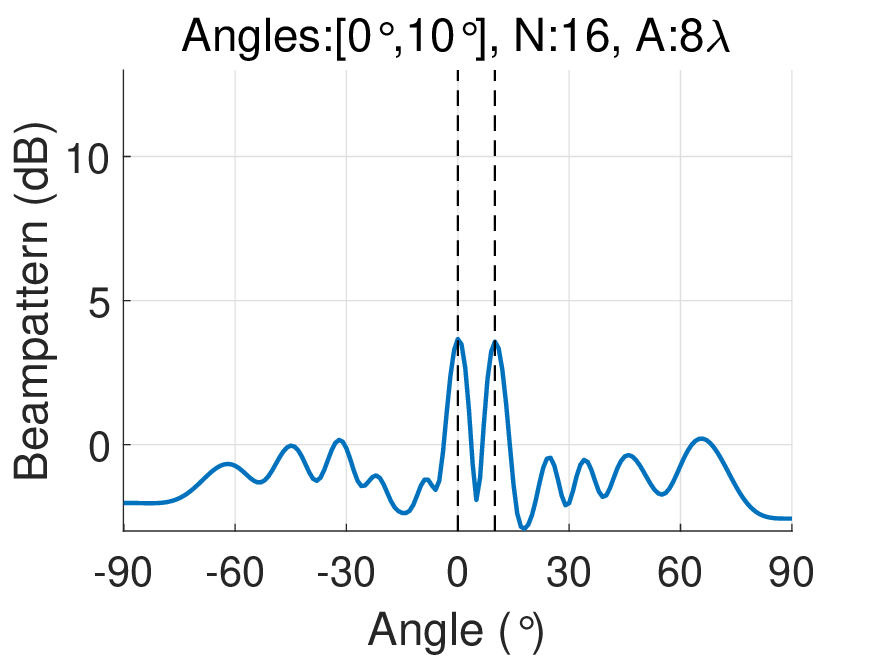}}	
	\centering
	\subfigure[]{
		\includegraphics[width=5.2cm]{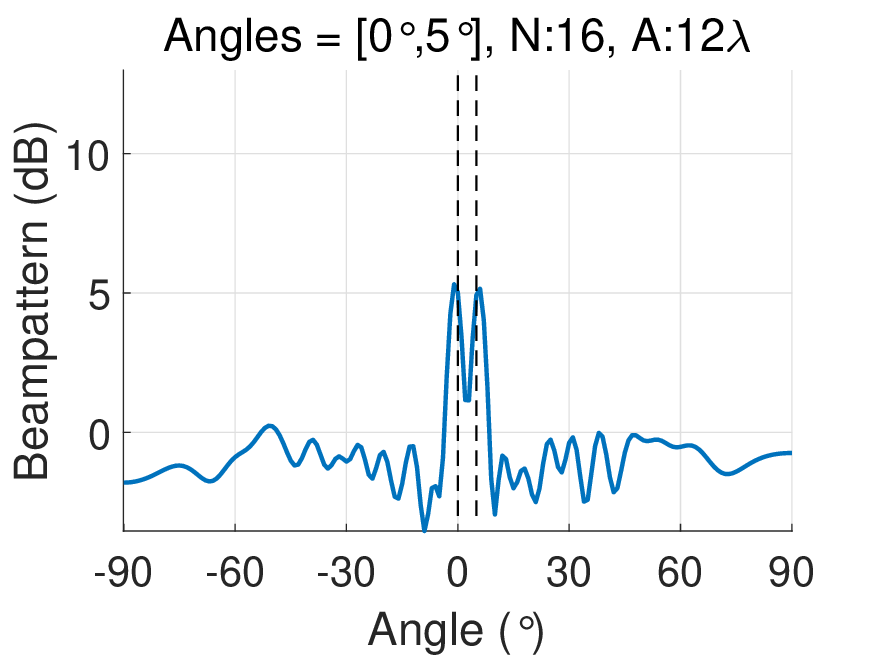}}
	\centering
	\subfigure[]{
		\includegraphics[width=5.2cm]{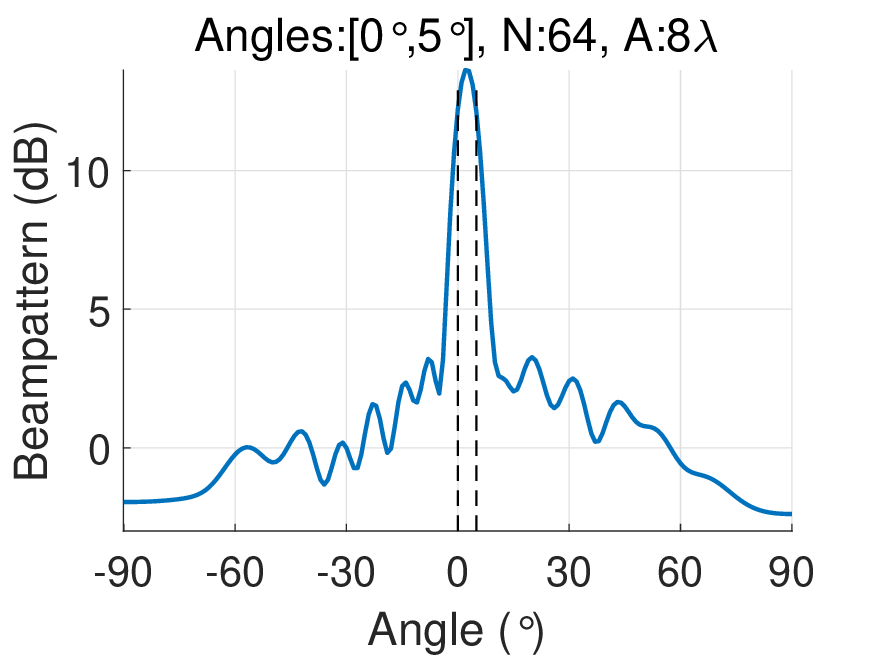}}
        \centering
        \subfigure[]{
		\includegraphics[width=5.2cm]{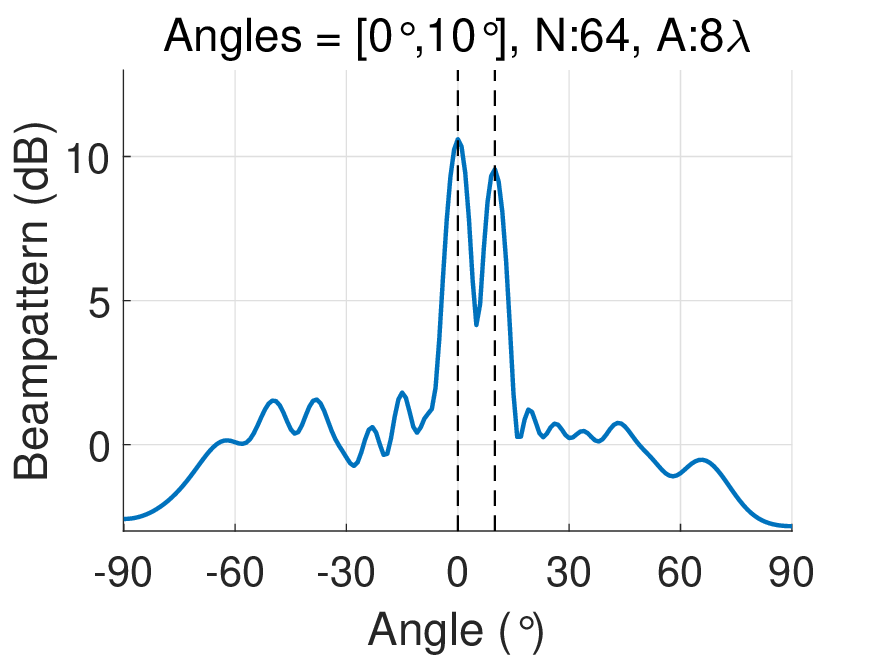}}
        \centering
        \subfigure[]{
		\includegraphics[width=5.2cm]{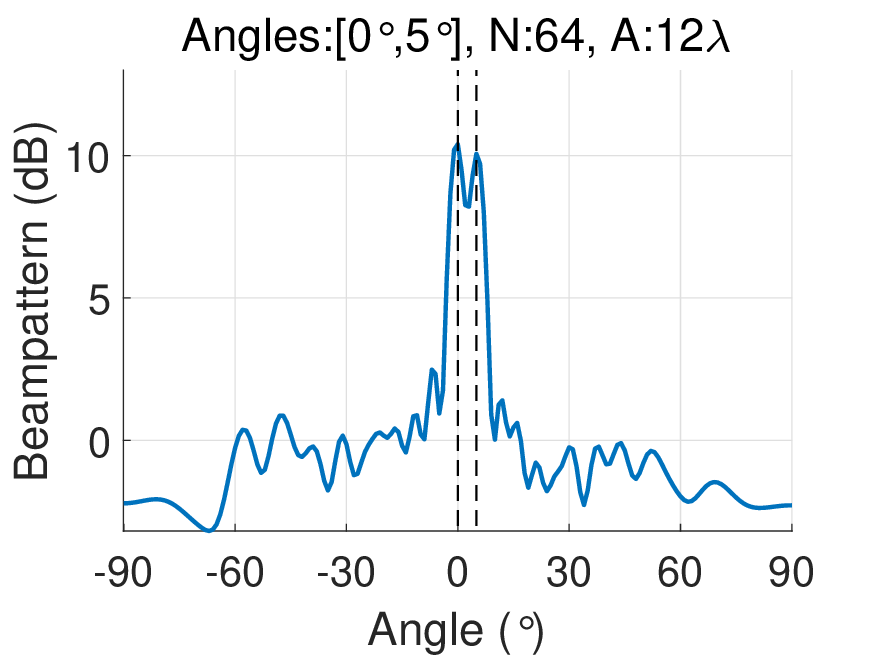}}
	\caption{Beampatterns (dB) of different system configurations: (a) Angles Separation: $5^\circ$, Element number: $16$, Size: $8\lambda$. (b) Angles Separation: $10^\circ$, Element number: $16$, Size: $8\lambda$. (c) Angles Separation: $5^\circ$, Element number: $16$, Size: $12\lambda$. (d) Angles Separation: $5^\circ$, Element number: $64$, Size: $8\lambda$. (e) Angles Separation: $10^\circ$, Element number: $64$, Size: $8\lambda$. (f) Angles Separation: $5^\circ$, Element number: $64$, Size: $12\lambda$. }
	\label{angle_separation}		
\end{figure*}

\begin{table}[t]
\centering
\caption{{Confirmation of the joint effect of position and phase optimization of fRIS}}
\scalebox{1}{
\begin{tabular}{|c|c|c|c|}
\hline
  & {Objective function}   & {Commun.}  & {Sensing}   \\
 & Value  &  MSE &MSE\\
\hline  
Joint  & \multirow{2}*{0.27}  & \multirow{2}*{0.23} & \multirow{2}*{0.31}\\
optimization &   &  &\\
\hline 
Position & \multirow{2}*{0.79}  & \multirow{2}*{1.1} & \multirow{2}*{0.48}\\
only &  &  & \\
\hline
 Phase shift & \multirow{2}*{0.57}  & \multirow{2}*{0.67}& \multirow{2}*{0.47}\\
only &  &  & \\
\hline 
No phase shift or & \multirow{2}*{2.07}  & \multirow{2}*{3.51}& \multirow{2}*{0.63}\\
position optimization &  &  & \\
\hline 
\end{tabular}}
\label{DoFs_validation}
\end{table}

In Fig. \ref{fig:convergence}{(a)}, we examine the convergence performance of the proposed algorithm for the fRIS-aided ISAC system and compare it with the benchmarks. As expected, the proposed approach demonstrates rapid convergence, stabilizing after $20$ iterations. Besides,  the MSE of the proposed scheme outperforms the baselines under the same element number, while yielding comparable performance as the ``Conven (HWL)'' case. This is because the continuous and flexible element position optimization provides additional spatial DoFs. {In addition, the DoFs of element position optimization are validated in Table \ref{DoFs_validation}. As compared to position-only/phase-only optimization, the joint optimization outperforms the individual optimization schemes, implying that repositioning does not overlap with the existing optimization dimension.} {To assess the impact of initialization, we conduct $100$ trials under a fixed channel realization and different initialization. As shown in Fig.~\ref{fig:convergence}{(b)}, the convergence exhibits small variation, with a mean of $0.1461$, a variance of $0.0065$, and a range of [$0.0577$, $0.4922$]. This indicates good robustness and stable convergence behavior.}
\begin{table}[!t]
\caption{Comparisons in terms of Average Performance Value \\and Time Complexity per iteration}
    \label{running_time}
    \centering
    \scalebox{1}{
    \begin{tabular}{|c|c|c|}
    \hline
    \multirow{2}*{Scheme} & \multirow{2}*{Objective Function Value}  & {Elapsed Time} \\
    &  & (s/iteration)  \\
    \hline
    Proposed: & \multirow{2}*{0.27}  & \multirow{2}*{24.6} \\
     $N = 25$ &  &  \\
    \hline
    Conven (SameNum): & \multirow{2}*{0.57}  & \multirow{2}*{1.33} \\
      $N = 25$ &  &  \\
    \hline
    Conven (HWL):  & \multirow{2}*{0.32}  & \multirow{2}*{47.75} \\
      $N = 81$ &  &  \\
    \hline
    \end{tabular}}
    \end{table}

 To further validate the advantages of fRIS, we compare the average objective function value and the time complexity of the proposed scheme with those of the conventional RIS-aided schemes in Table \ref{running_time}. Specifically, in the ``Conven (SameNum)'' scheme, the conventional RIS has the same element number as fRIS ($N=25$), while in the ``Conven (HWL)'' scheme, the elements have fixed half-wavelength spacing under the same aperture, where the element number is $N=81$. It is seen that the proposed scheme achieves better results compared to ``Conven (SameNum)'' due to additional DoFs. When compared with the ``Conven (HWL)'' scheme, the proposed scheme exhibits comparable performance but significantly less elapsed time per iteration. These results demonstrate that employing the fRIS can compensate for performance losses caused by reducing RIS elements, and alleviate the complexity of high-dimensional optimization.

For the sensing performance evaluation, we demonstrate the fRIS beampattern under different weighting coefficients ($\alpha=0.1, 0.5,0.9$), and compare with the ideal case and the reference beampattern in Figs. \ref{Fig_Beam_01}-\ref{Fig_Beam_09}.  It is observed that beams are consistently formed in the target directions. In the case of $\alpha=0.1$ where the system is biased towards communication, the beam is not as similar as the sensing-only case and exhibits higher sidelobe levels. In contrast, when $\alpha=0.5,0.9$ where the system is increasingly biased towards the sensing, the formed beams gradually align with the reference beampattern. {While a discrepancy in beampattern energy exists between the “Proposed” and “Sensing-only” cases, it arises from the joint optimization process, where the inherent tradeoff introduces a mismatch between the covariance of the actual signal $\mathbf{v}$ and that of the reference signal $\mathbf{s}_r$. Consequently, the realized signal may differ from $\mathbf{s}_r$ in both shape and power. In addition, since $\mathbf{s}_r$ is a virtual design target and not physically transmitted, this mismatch does not violate energy conservation. } On the other hand, Fig. \ref{Fig_Beam_Comp} compares the beampattern of different approaches. It is observed that all the beampatterns have formed peaks in the target directions. Meanwhile, the proposed method achieves lower sidelobe levels compared to the conventional case and higher peaks compared to the ``DPS" cases.  

{To further evaluate the sidelobe levels and the energy leakage, we demonstrate integrated sidelobe to mainlobe ratio (ISMR) \cite{ISMR} in Table \ref{tab:ISMR}. Specifically, ISMR is computed as the ratio of the integrated power in the sidelobe region to that in the mainlobe region. Given the target directions $\left[-20^\circ, 5^\circ, 30^\circ\right]$, we set the mainlobe regions as $\left[-30^\circ,-10^\circ\right] \cup \left[-5^\circ,15^\circ\right] \cup \left[20^\circ,40^\circ\right]$.} {It is observed that higher $\alpha$ yields lower ISMR, demonstrating better sidelobe suppression, while the proposed fRIS-aided design consistently achieves lower ISMR than the conventional RIS scheme with the same amount of element. For further sidelobe suppression, we may consider incorporating ISMR into optimization. Such enhancement is non-trivial and worth exploring in future work.} 

{Under closely spaced targets, we simulate the beampatterns when two targets are placed at $0^\circ$ and another at $5^\circ$ or $10^\circ$. The aperture size of fRIS is set as $A \! \in \! \{8\lambda, 12\lambda\}$ and the number of elements is set as $N \!\in\! \{16, 64\}$. Fig. \ref{angle_separation} reveals that larger angular separation improves beam distinguishability, while increasing aperture $A$ narrows the mainlobes. For instance, while $A \!=\! 8\lambda$ leads to merged beams at $5^\circ$ separation, $A\! = \!12\lambda$ achieves separated peaks. Additionally, increasing $N$ can suppress the sidelobe and enhance the mainlobe gain. }

\begin{table}[t]
    \centering
    \caption{ISMR under different settings.}
   \begin{tabular}{|c|c|c|c|}
    \hline
         & $\alpha=0.1$ & $\alpha=0.5$ (Default) & $\alpha=0.9$ \\
         \hline
        ISMR (dB) & 1.49 & -0.22 & -0.51\\
        \hline
         & Conven (SameNum):  & Conven (HWL):   & \multirow{2}*{DPS} \\
        &   $N = 25$ &   $N = 81$ &   \\
        \hline
        ISMR (dB) & 2.65 & -0.58 & -0.18 \\
        \hline
    \end{tabular} 
    \label{tab:ISMR}
\end{table}

\begin{figure}[!t] 
    \includegraphics[scale=0.48]{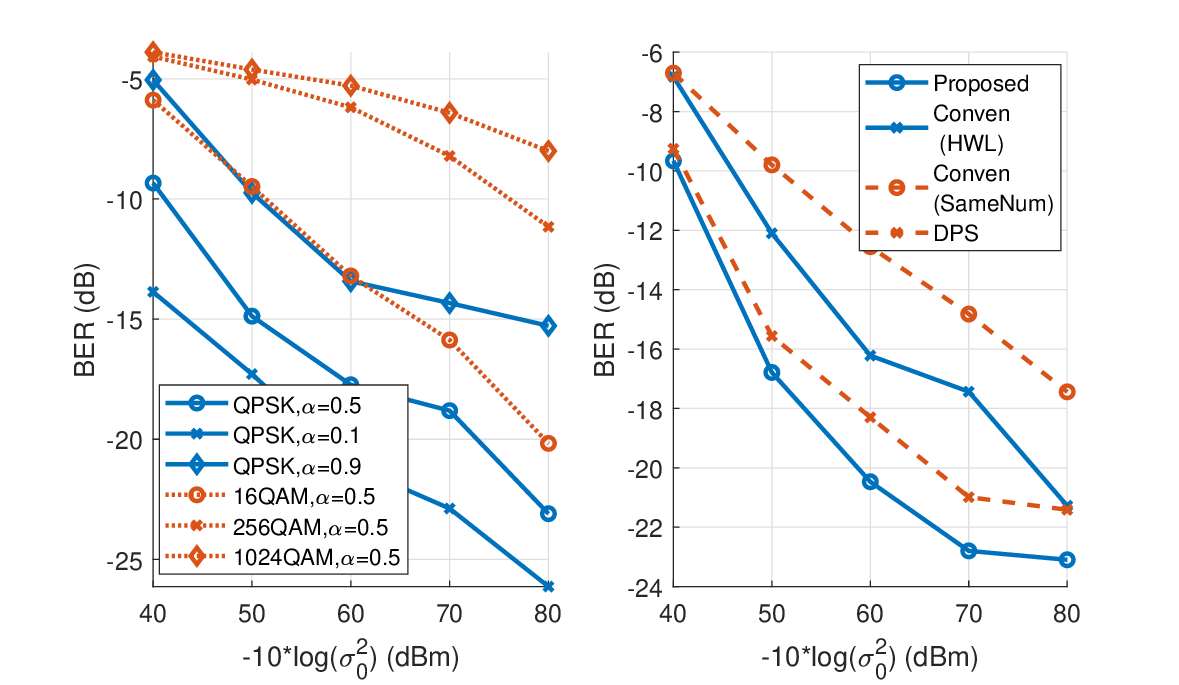}    \centerline{\footnotesize{(a)}~~~~~~~~~~~~~~~~~~~~~~~~~~~~~~~~~~~~ (b)}  
    \caption{(a) Average BER (dB) versus different receive noise power (dBm) under different modulations and weighting coefficients. (b) Average BER (dB) versus different receive noise power (dBm) with different approaches.}
    \label{fig:BER}
\end{figure}

\begin{figure}[!t]
    % \centering
    \includegraphics[scale=0.53]{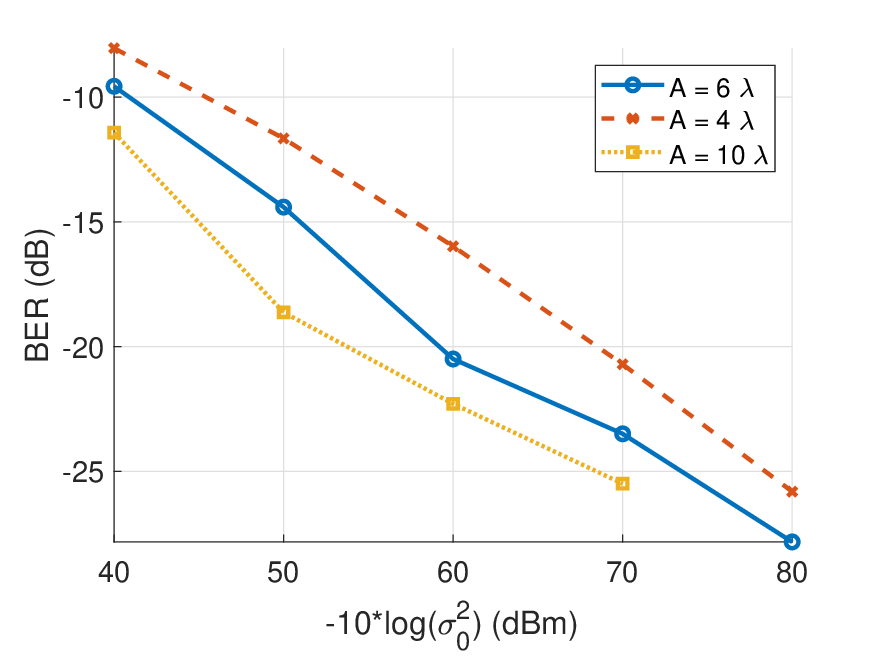}
    \caption{Average BER (dB) versus different receive noise power (dBm) under different fRIS sizes.}
    \label{fig:BER-Aper}
\end{figure}

% \begin{figure}[!t]
%     \centering
%     \includegraphics[scale=0.43]{BER_alpha_Mod.eps}
%     \caption{Average BER (dB) versus different receive noise power (dBm) under different modulations and weighting coefficients.}
%     \label{fig:BER}
% \end{figure}
% \begin{figure}[!t]
%     \centering
%     \includegraphics[scale=0.43]{BER_algo.eps}
%     \caption{Average BER (dB) versus different receive noise power (dBm) with different approaches.}
%     \label{fig:BER_comp}
% \end{figure}
In addition, we evaluate the communication bit error rate (BER) of the system under different settings. Fig. \ref{fig:BER}{(a)} depicts the BER of the proposed scheme versus various noise power levels. As expected, the average BER decreases when the noise power decreases, and QPSK modulation exhibits lower BER when compared to higher-order modulations. Moreover, when $\alpha$ decreases, the BER improves as the system becomes more communication-biased. On the other hand, Fig. \ref{fig:BER}{(b)} compares the average BER of the proposed method to the benchmarks. As shown, all cases exhibit a decreasing BER trend as noise power level reduces, while the proposed method consistently achieves a lower BER compared to the baselines. In Fig. \ref{fig:BER-Aper}, the impact of the fRIS size on BER is evaluated, where we find that the larger fRIS aperture size yields lower BER. This is because increasing $A$ can form narrower beams, which increase the received energy for users.

\section{Conclusions} \label{conclusions}
This paper proposed a fRIS-aided ISAC system to achieve multi-user communication and multi-target sensing, where a joint communication MSE and sensing beampattern mismatch minimization problem was formulated. To address the problem, an AM-based algorithm was developed, which involved various techniques.  Simulation results validated that the fRIS significantly reduces the sensing beampattern mismatch and communication BER. {Future research may explore new fRIS architectures in ISAC systems, such as active or beyond diagonal fRIS designs. In addition, it is interesting to study the modeling of fRIS under rich-scattering environments.}

\begin{appendices}   
\section{Derivation of the function $\widetilde{f}_0$ in (\ref{p_obj1})}  \label{app_a}

To simplify (\ref{optimize_p}{a}), we first expand the two square norms, which is given by
\begin{align} \label{f0_expand}
    &\!\!\!\alpha \cdot \left(\mathbf{x}^\mathrm{H} \mathbf{G}^\mathrm{H} \boldsymbol{\Theta} 
   \boldsymbol{\Theta}^\mathrm{H} \mathbf{G} \mathbf{x} - 2 \mathfrak{R}\left\lbrace\mathbf{s}_r^\mathrm{H} \boldsymbol{\Theta}^\mathrm{H} \mathbf{G} \mathbf{x}\right\rbrace \right)\!+\! (1\!-\! \alpha) \cdot  \notag \\
   & \!\!\!\!\left(\omega^2\mathbf{x}^\mathrm{H}  \mathbf{G}^\mathrm{H}  \boldsymbol{\Theta}  \mathbf{H}_{rc}\mathbf{H}_{rc}^{\mathrm{H}} \boldsymbol{\Theta}^\mathrm{H} \mathbf{G} \mathbf{x} \!\!-\!\! 2 \omega \mathfrak{R}\left \lbrace \mathbf{s}_c^{\mathrm{H}} \mathbf{H}_{rc}^{\mathrm{H}} \boldsymbol{\Theta}^\mathrm{H} \mathbf{G}\mathbf{x} \right\rbrace\right).
\end{align}
Here, $\mathbf{s}_r^\mathrm{H}\mathbf{s}_r$ and $\mathbf{s}_c^\mathrm{H}\mathbf{s}_c$ are constants, which have been omitted. 

Observing the  term $\mathbf{x}^\mathrm{H} \mathbf{G}^\mathrm{H} \boldsymbol{\Theta} 
   \boldsymbol{\Theta}^\mathrm{H} \mathbf{G} \mathbf{x}$, it is unrelated to  $\mathbf{p}$ due to the following reason. Given  $\boldsymbol{\Theta}\boldsymbol{\Theta}^\mathrm{H}=\mathbf{I}$ and the expression of $\mathbf{G}$ in (\ref{chan_G}), this term can be written as $\zeta_G \mathbf{x}^\mathrm{H}\mathbf{a}_t  \mathbf{a}_r^\mathrm{H}(\mathbf{p})  \mathbf{a}_r(\mathbf{p})  \mathbf{a}_t^{\mathrm{H}}  \mathbf{x}$.
   Since $\mathbf{a}_r(\mathbf{p})$ has the structure of array manifold of (\ref{FA_steering_vector}), $\mathbf{a}_r^\mathrm{H}(\mathbf{p})  \mathbf{a}_r(\mathbf{p}) = N $ always holds regardless of $\mathbf{p}$. Therefore, this term becomes $\zeta_G  N \mathbf{x}^\mathrm{H}\mathbf{a}_t  \mathbf{a}_t^{\mathrm{H}}  \mathbf{x}$, which is  unrelated to $\mathbf{p}$ and ignorable.

   For the term $-2 \mathfrak{R}\lbrace\mathbf{s}_r^\mathrm{H} \boldsymbol{\Theta}^\mathrm{H} \mathbf{G} \mathbf{x}\rbrace$, the expression of $\mathbf{G}$ can be substituted, yielding $-2 \sqrt{\zeta_G} \mathfrak{R}\lbrace\mathbf{s}_r^\mathrm{H} \boldsymbol{\Theta}^\mathrm{H}  \mathbf{a}_r(\mathbf{p})  \mathbf{a}_t^{\mathrm{H}} \mathbf{x}\rbrace$. Leveraging the cyclic property of trace, it is re-expressed as $-2 \sqrt{\zeta_G} \mathfrak{R}\lbrace \widetilde{\mathbf{a}}_t^\mathrm{H}  \mathbf{a}_r(\mathbf{p}) \rbrace$, where $\widetilde{\mathbf{a}}_t^\mathrm{H} = \mathbf{a}_t ^\mathrm{H} \mathbf{x}\mathbf{s}_r^\mathrm{H}\boldsymbol{\Theta}^\mathrm{H}$.

   Focusing on the third term, it is rewritten as
   \begin{subequations}
     \begin{align}
        &\omega^2  \mathbf{x}^\mathrm{H}  \mathbf{G}^\mathrm{H}  \boldsymbol{\Theta}  \mathbf{H}_{rc}\mathbf{H}_{rc}^{\mathrm{H}} \boldsymbol{\Theta}^\mathrm{H} \mathbf{G} \mathbf{x}  \\
           \overset{\mathrm{(a)}}{\longrightarrow} & c_0 \mathbf{a}_r^{\mathrm{H}}(\mathbf{p}) \boldsymbol{\Theta}  \mathbf{A}_{rc} (\mathbf{p})\boldsymbol{\Sigma}_{rc} \boldsymbol{\Sigma}_{rc}^{\mathrm{H}}  \mathbf{A}_{rc}^{\mathrm{H}} (\mathbf{p})\boldsymbol{\Theta}^\mathrm{H} \mathbf{a}_r(\mathbf{p})   \\
           \overset{\mathrm{(b)}}{\longrightarrow} & c_0 \sum_{k=1}^{K}  \zeta_k   \mathbf{a}_r^{\mathrm{H}}(\mathbf{p})   \boldsymbol{\Theta}   \mathbf{a}_{c,k}( \mathbf{p}) \mathbf{a}_{c,k}^\mathrm{H}(\mathbf{p})\boldsymbol{\Theta}^\mathrm{H} \mathbf{a}_r(\mathbf{p}) \\
           \overset{\mathrm{(c)}}{\longrightarrow} & c_0 \sum_{k=1}^{K}  \zeta_k   \mathbf{a}_r^{\mathrm{H}}(\mathbf{p})   \widetilde{\mathbf{A}}_{c,k}( \mathbf{p}) \boldsymbol{\theta}    \boldsymbol{\theta}^\mathrm{H} \widetilde{\mathbf{A}}_{c,k}^\mathrm{H}(\mathbf{p}) \mathbf{a}_r(\mathbf{p}).
    \end{align}
   \end{subequations}
   In  (a), the expression of $\mathbf{G}$ and $\mathbf{H}_{rc}$ are substituted, while $c_x = \mathbf{a}_t ^\mathrm{H}\mathbf{x}\mathbf{x}^{\mathrm{H}}\mathbf{a}_t$ and $c_0 = \omega^2 \zeta_G c_x$ incorporate the variable-independent terms. In (b), it is derived based on the matrix structure in (\ref{chan_Hrc_aff}{a}) and (\ref{chan_Hrc_aff}{b}). The operation (c) leverages the diagonal property of $\boldsymbol{\Theta}$, where we denote $\widetilde{\mathbf{A}}_{c,k}( \mathbf{p}) = \mathrm{Diag}(\mathbf{a}_{c,k}( \mathbf{p}))$.

    For the last term in (\ref{f0_expand}), it is transformed as follows:
    \begin{subequations}
        \begin{align}
       & - 2 \omega \mathfrak{R}\left \lbrace \mathbf{s}_c^{\mathrm{H}} \mathbf{H}_{rc}^{\mathrm{H}} \boldsymbol{\Theta}^\mathrm{H} \mathbf{G}\mathbf{x} \right\rbrace \\
        \overset{\mathrm{(a)}}{\longrightarrow} & - 2 \omega \sqrt{\zeta_G} \mathfrak{R}\left \lbrace \mathbf{s}_c^{\mathrm{H}}  \boldsymbol{\Sigma}_{rc}^{\mathrm{H}} \mathbf{A}_{rc}^{\mathrm{H}} (\mathbf{p}) 
 \boldsymbol{\Theta}^\mathrm{H}  \mathbf{a}_r(\mathbf{p})  \mathbf{a}_t^{\mathrm{H}} \mathbf{x} \right\rbrace \\
 \overset{\mathrm{(b)}}{\longrightarrow} &- 2 \omega \sqrt{\zeta_G} \mathfrak{R}\left \lbrace  \mathrm{tr}\left(\mathbf{S} \mathbf{A}_{rc}^{\mathrm{H}} (\mathbf{p}) 
  \widetilde{\mathbf{A}}_{r}( \mathbf{p})  \right)\right\rbrace,
   \end{align}
    \end{subequations}
   where (a) is the substitution operation, while (b) utilizes the cyclic property of trace and the diagonal property of $\boldsymbol{\Theta}$. In addition, we denote $\widetilde{\mathbf{A}}_{r}( \mathbf{p}) = \mathrm{Diag}(\mathbf{a}_{r}( \mathbf{p}))$ and $\mathbf{S}=\boldsymbol{\theta}^* \mathbf{a}_t^{\mathrm{H}} \mathbf{x} \mathbf{s}_c^{\mathrm{H}}  \boldsymbol{\Sigma}_{rc}^{\mathrm{H}}$.

   By integrating all the simplified terms, the objective function can be expressed as (\ref{p_obj1}).  This ends the derivation.

   \section{Derivation of the function $\widetilde{f}_1$ in (\ref{p_obj2})}  \label{app_b}
    For (\ref{p_obj1}), it can be simplified if we consider the position optimization of the $n$-th element   $\mathbf{p}_n$ while fixing others. First, examining the first term $-2 \alpha \sqrt{\zeta_G} \mathfrak{R} \lbrace \widetilde{\mathbf{a}}_t^\mathrm{H}  \mathbf{a}_r(\mathbf{p}) \rbrace$, it is the sum of $N$ number, which is expressed as
    \begin{subequations}
    \begin{align}
        & \sum_{n=1}^{N} -2 \alpha\sqrt{\zeta_G} \mathfrak{R}\left\lbrace [\widetilde{\mathbf{a}}_t^\mathrm{H} ]_n e^{\jmath \frac{2\pi}{\lambda}d_{r,n}(\mathbf{p}_n) } \right\rbrace \\
         \overset{\mathrm{(a)}}{\longrightarrow} &-2 \alpha \sqrt{\zeta_G} \mathfrak{R}\left\lbrace [\widetilde{\mathbf{a}}_t^\mathrm{H} ]_n e^{\jmath \frac{2\pi}{\lambda}d_{r,n}(\mathbf{p}_n) } \right\rbrace \\
         {\longrightarrow} &-\nu_{n}  \cos \left( \xi_n +\frac{2\pi}{\lambda}d_{r,n}(\mathbf{p}_n) \right). 
    \end{align}        
    \end{subequations}
    In particular, $[\widetilde{\mathbf{a}}_t^\mathrm{H} ]_n$ is the $n$-th entry of $\widetilde{\mathbf{a}}_t^\mathrm{H}$, while $\rho_n=|[\widetilde{\mathbf{a}}_t^\mathrm{H} ]_n|$
 and $\xi_n = \angle  ([\widetilde{\mathbf{a}}_t^\mathrm{H} ]_n  )$. Additionally, $\nu_{n}= 2\alpha  \sqrt{\zeta_G} \rho_n$ is a variable-independent value. The symbol $d_{r,n}(\mathbf{p}_n)$ is determined by $\mathbf{p}_n$, i.e., $p_{x,n} \sin (\phi_r)\cos (\psi_r) + p_{y,n} \sin (\psi_r)$. Moreover, in (a), only the term related to $\mathbf{p}_n$ is kept since other element positions are considered constant.

For the second term, it can be rewritten in an element-wise manner, i.e.,
 \begin{subequations} \label{temp_appB_2}
    \begin{align}
    &(1\!-\! \alpha) c_0 \sum_{k=1}^{K}  \zeta_k   \mathbf{a}_r^{\mathrm{H}}(\mathbf{p})   \widetilde{\mathbf{A}}_{c,k}( \mathbf{p}) \boldsymbol{\theta}    \boldsymbol{\theta}^\mathrm{H} \widetilde{\mathbf{A}}_{c,k}^\mathrm{H}(\mathbf{p}) \mathbf{a}_r(\mathbf{p})   \\
    \longrightarrow & (1\!-\! \alpha) c_0 \sum_{k=1}^{K} \sum_{i=1}^{N} \sum_{j=1}^{N}  \zeta_k \left[\boldsymbol{\theta}    \boldsymbol{\theta}^\mathrm{H}\right]_{i,j} \!\!\!\cdot \left[ \mathbf{a}_r^{\mathrm{H}}(\mathbf{p})   \widetilde{\mathbf{A}}_{c,k}( \mathbf{p})\right]_i  \notag\\
    & \quad \quad \quad \quad \quad \quad \quad \quad \quad \quad \quad \quad \cdot \left[ \widetilde{\mathbf{A}}_{c,k}^\mathrm{H}(\mathbf{p}) \mathbf{a}_r(\mathbf{p})\right]_j \\
     \longrightarrow & (1\!-\! \alpha) c_0 \sum_{k=1}^{K} \sum_{i=1}^{N} \sum_{j=1}^{N}  \zeta_k \left[\boldsymbol{\theta}    \boldsymbol{\theta}^\mathrm{H}\right]_{i,j} \!\!\!  \notag\\ 
     & \! \! \cdot e^{-\jmath \frac{2\pi}{\lambda} \left(d_{r,i}(\mathbf{p}_i)-d_{c,k,i}(\mathbf{p}_i)\right)} 
    \!\cdot\! e^{\jmath \frac{2\pi}{\lambda} \left(d_{r,j}(\mathbf{p}_j)-d_{c,k,j}(\mathbf{p}_j)\right)},
 \end{align}  
 \end{subequations}
 where $d_{c,k,n}(\mathbf{p}_n)=p_{x,n} \sin (\phi_{c,k})\cos (\psi_{c,k}) + p_{y,n} \sin (\psi_{c,k})$. 
When $i=j$, the complex exponential terms are mutually canceled out and become constants unrelated to $\mathbf{p}$. Besides,  the terms of $i \neq j$ appear in conjugate pairs since $\boldsymbol{\theta} \boldsymbol{\theta}^\mathrm{H}$ is a Hermitian matrix,  Owing to this characteristics, when considering the $n$-th element (i.e. $i=n\text{ or }j=n$) and ignoring the constant terms of $i=j$,  (\ref{temp_appB_2}{c}) is  reduced to
\begin{align} \label{temp3}
 & (1\!-\! \alpha) c_0 \sum_{k=1}^{K} \zeta_k \sum_{i=1,i\neq n}^{N} 2 \mathfrak{R} \left\lbrace \left[\boldsymbol{\theta}    \boldsymbol{\theta}^\mathrm{H}\right]_{i,n} \right.\!\!\!  \notag\\ 
     & \! \! \left.\cdot e^{-\jmath \frac{2\pi}{\lambda} \left(d_{r,i}(\mathbf{p}_i)-d_{c,k,i}(\mathbf{p}_i)\right)} 
    \cdot e^{\jmath \frac{2\pi}{\lambda} \left(d_{r,n}(\mathbf{p}_n)-d_{c,k,n}(\mathbf{p}_n)\right)}\right\rbrace.
 \end{align} 
 Notably, when fixing the positions of other elements, the term $[\boldsymbol{\theta}    \boldsymbol{\theta}^\mathrm{H}]_{i,n} \cdot e^{-\jmath \frac{2\pi}{\lambda} \left(d_{r,i}(\mathbf{p}_i)-d_{c,k,i}(\mathbf{p}_i)\right)}$ is unrelated to $\mathbf{p}_n$, which can be viewed constant as $\widetilde{c}_{i,n,k}$. We denote the phase of $\widetilde{c}_{i,n,k}$ as $\widetilde{\xi}_{i,n,k}=\angle \widetilde{c}_{i,n,k}$ while $\widetilde{c}_{i,n,k}$ has an amplitude of $1$. As a result, (\ref{temp3}) can be written as
 \begin{subequations}
   \begin{align}
      &  \!\!\!\!\!\!\!\!\!\!\!\!\!\!\!2(1\!-\! \alpha) c_0 \sum_{k=1}^{K} \!\zeta_k \!\!\!\!\sum_{i=1,i\neq n}^{N}\!\!\!\! \! \mathfrak{R} \left\lbrace \widetilde{c}_{i,n,k} e^{\jmath \frac{2\pi}{\lambda} \left(d_{r,n}(\mathbf{p}_n)-d_{c,k,n}(\mathbf{p}_n)\right)}\right\rbrace\\
      \longrightarrow &  \sum_{k=1}^{K}  \! \sum_{i=1,i\neq n}^{N}\!\!  \widetilde{\nu} \zeta_k\cos \left( \widetilde{\xi}_{i,n,k} +  \frac{2\pi}{\lambda} \Delta \widetilde{d}_{k,n} (\mathbf{p}_n) \right),
 \end{align}   
 \end{subequations}
 where we denote $\Delta \widetilde{d}_{k,n}(\mathbf{p}_n) =d_{r,n}(\mathbf{p}_n)-d_{c,k,n}(\mathbf{p}_n)$ and $\widetilde{\nu}=2(1\!-\! \alpha) c_0 $ for simplicity.

Focusing on the last term, it can be transformed as follows:
\begin{subequations}
  \begin{align}
   & - 2 (1\!-\! \alpha) \omega \sqrt{\zeta_G} \mathfrak{R}\left \lbrace  \mathrm{tr}\left(\mathbf{S} \mathbf{A}_{rc}^{\mathrm{H}} (\mathbf{p}) 
  \widetilde{\mathbf{A}}_{r}( \mathbf{p})  \right)\right\rbrace \\
  \longrightarrow & \!\sum_{n=1}^{N}\sum_{k=1}^{K}  \! \!- \overline{\nu} \mathfrak{R}\left \lbrace  s_{n,k} e^{\jmath \frac{2\pi}{\lambda}(d_{r,n}(\mathbf{p}_n)-d_{c,k,n}(\mathbf{p}_n) ) }  \right\rbrace \\
  \longrightarrow & \sum_{k=1}^{K}  - \overline{\nu} \mathfrak{R}\left \lbrace  s_{n,k} e^{\jmath \frac{2\pi}{\lambda}\Delta \widetilde{d}_{k,n}(\mathbf{p}_n)) }  \right\rbrace \\
  \longrightarrow & \sum_{k=1}^{K} \! -  \overline{\nu} ~\overline{\rho}_{n,k} \cos\!\left (  \overline{\xi}_{n,k} \!+\! \frac{2\pi}{\lambda}\Delta \widetilde{d}_{k,n}(\mathbf{p}_n)  \right),
\end{align}  
\end{subequations}
where $s_{n,k}=[\mathbf{S}]_{n,k}$ with the amplitude of $\overline{\rho}_{n,k}= |s_{n,k}|$ and the phase of $\overline{\xi}_{n,k}= \angle s_{n,k}$. We also denote $\overline{\nu}=2 (1\!-\! \alpha) \omega \sqrt{\zeta_G} $ for brevity. 

This completes the derivation.

 \section{Gradient and Hessian matrix Derivation of $\widetilde{f}_1$}  \label{app_c}
Given that $\widetilde{f}_1(\mathbf{p}_n)$ is the sum of the sub-functions $\widetilde{f}_{1,1}$, $\widetilde{f}_{1,2}$, $\widetilde{f}_{1,3}$ shown in (\ref{p_obj2}), the derivative of $\widetilde{f}_1(\mathbf{p}_n)$ is equal to the respective derivatives of the sub-functions. 

For the sub-function $\widetilde{f}_{1,1}$, the first-order derivative w.r.t. $p_{x,n}$ and $p_{y,n}$ is the derivative of a composite function, where the outer function is a cosine function while the interior function is linear. Therefore, we can adopt the chain rule of derivatives. First, we derive the the first-order derivatives of $\widetilde{f}_{1,1}$:
\begin{subequations} \label{first_deri_1}
   \begin{align}
    &\left[\!\frac{\partial \widetilde{f}_{1,1}}{\partial p_{x,n}}, \frac{\partial \widetilde{f}_{1,1}}{\partial p_{y,n}}\right]^\mathrm{T}\!=\! \nu_{n}  \frac{2\pi}{\lambda} \boldsymbol{\kappa}_{r} \sin \! \left( \tau_{n} \right), 
\end{align} 
\end{subequations}
where $\boldsymbol{\kappa}_{r}=[\kappa_{r,x},\kappa_{r,x}]^\mathrm{T}= [\sin (\phi_r)\cos (\psi_r),\sin (\psi_r)]^\mathrm{T}$ and $\tau_{n}=\xi_n \!+ \! \frac{2\pi}{\lambda}d_{r,n}$. Based on (\ref{first_deri_1}), the second-order derivatives are derived as
   \begin{align}\label{second_deri_1}
    &\begin{bmatrix}
        \frac{\partial^2 \widetilde{f}_{1,1}}{\partial p_{x,n}^2} & \frac{\partial^2 \widetilde{f}_{1,1}}{\partial p_{x,n} \partial p_{y,n}}\\
        \frac{\partial^2 \widetilde{f}_{1,1}}{\partial p_{y,n} \partial p_{x,n}} & \frac{\partial^2 \widetilde{f}_{1,1}}{\partial p_{y,n}^2}
    \end{bmatrix} = \nu_{n} \frac{4\pi^2}{\lambda^2} \cos \! \left( \tau_{n}\right) \boldsymbol{\kappa}_{r} \boldsymbol{\kappa}_{r}^\mathrm{T}.
\end{align}

%   \frac{\partial^2 \widetilde{f}_{1,1}}{\partial p_{x,n}^2} \!&=\! \nu_{n}  \frac{4\pi^2}{\lambda^2} \kappa_{r,x}^2 \cos \! \left( \xi_n \!+ \! \frac{2\pi}{\lambda}d_{r,n} \!\right), \\ \frac{\partial^2 \widetilde{f}_{1,1}}{\partial p_{x,n} \partial p_{y,n}} \!&=\! \frac{\partial^2 \widetilde{f}_{1,1}}{\partial p_{y,n} \partial p_{x,n}} \notag \\ & = \nu_{n}   \frac{4\pi^2}{\lambda^2} \kappa_{r,x} \kappa_{r,y} \cos \! \left( \xi_n \!+ \! \frac{2\pi}{\lambda}d_{r,n} \!\right), \\ \frac{\partial^2 \widetilde{f}_{1,1}}{\partial p_{y,n}^2} \!&=\! \nu_{n}  \frac{4\pi^2}{\lambda^2} \kappa_{r,y}^2 \cos \! \left( \xi_n \!+ \! \frac{2\pi}{\lambda}d_{r,n} \!\right).
 For the second sub-function $\widetilde{f}_{1,2}$, the first-order derivatives and the second-order derivatives can be derived as
\begin{subequations} \label{first_deri_2}
   \begin{align}
   &\!\!\!\left[\frac{\partial \widetilde{f}_{1,2}}{\partial p_{x,n}},  
\frac{\partial \widetilde{f}_{1,2}}{\partial p_{y,n}}\right]=\sum_{i\neq n}^{N}\sum_{k=1}^{K} \!\! - \widetilde{\nu}_{k} \frac{2\pi}{\lambda}  \Delta \boldsymbol{\widetilde{\kappa}}_{k}  \sin \!\left( \widetilde{\tau}_{i,k,n}\right)\\
&\!\!\!\begin{bmatrix}
        \!\!\!\!\frac{\partial^2 \widetilde{f}_{1,2}}{\partial p_{x,n}^2} &\!\!\!\! \!\!\!\frac{\partial^2 \widetilde{f}_{1,2}}{\partial p_{x,n} \partial p_{y,n}}\\
        \frac{\partial^2 \widetilde{f}_{1,2}}{\partial p_{y,n} \partial p_{x,n}} &\!\! \!\!\!\!\frac{\partial^2 \widetilde{f}_{1,2}}{\partial p_{y,n}^2}
    \end{bmatrix}\!=\!\! \sum_{i\neq n}^{N}\sum_{k=1}^{K} \! - \widetilde{\nu}_{k} \frac{4\pi^2}{\lambda^2}  \Delta \mathbf{K}_k\cos \left( \widetilde{\tau}_{i,k,n}\right)\!, 
\end{align} 
\end{subequations}
where $\widetilde{\tau}_{i,k,n}=\!\widetilde{\xi}_{i,n} \!+\!  \frac{2\pi}{\lambda} \Delta \widetilde{d}_{k,n} $, $\Delta \mathbf{K}_k=\Delta \boldsymbol{\widetilde{\kappa}}_{k} \Delta \boldsymbol{\widetilde{\kappa}}_{k}^\mathrm{T}$, $\Delta \boldsymbol{\widetilde{\kappa}}_{k}=\boldsymbol{\kappa}_{r}-\boldsymbol{\kappa}_{c,k}$ and  $\boldsymbol{\kappa}_{c,k} =[\sin (\phi_{c,k})\cos (\psi_{c,k}),\sin (\psi_{c,k})]^\mathrm{T}$.

Similarly, the first-order and second-order derivatives of $\widetilde{f}_{1,3}$ can be given by
\begin{subequations} \label{first_deri_3}
   \begin{align}
    &\!\!\!\!\!\left[\frac{\partial \widetilde{f}_{1,3}}{\partial p_{x,n}},  
\frac{\partial \widetilde{f}_{1,3}}{\partial p_{y,n}}\right]= \sum_{k=1}^{K}  \overline{\nu}_{n,k}\frac{2\pi}{\lambda}  \Delta \boldsymbol{\widetilde{\kappa}}_{k} \sin\!\left ( \overline{\tau}_{n,k}\right)\\
&\!\!\!\!\!\begin{bmatrix}
        \!\!\!\frac{\partial^2 \widetilde{f}_{1,3}}{\partial p_{x,n}^2} & \!\!\!\!\frac{\partial^2 \widetilde{f}_{1,3}}{\partial p_{x,n} \partial p_{y,n}}\\
        \frac{\partial^2 \widetilde{f}_{1,3}}{\partial p_{y,n} \partial p_{x,n}} & \!\!\!\!\!\frac{\partial^2 \widetilde{f}_{1,3}}{\partial p_{y,n}^2}
    \end{bmatrix}\! \!=\!\! \sum_{k=1}^{K}  \overline{\nu}_{n,k}\frac{4\pi^2}{\lambda^2}   \Delta \mathbf{K}_k\cos\!\left ( \overline{\tau}_{n,k} \right)\!,
\end{align} 
\end{subequations}
where $\overline{\tau}_{n,k}= \overline{\xi}_{n,k} \!+\! \frac{2\pi}{\lambda}\Delta \widetilde{d}_{k,n}$.
%,\\
    % &\!\frac{\partial^2 \widetilde{f}_{1,3}}{\partial p_{x,n}^2}  \!=\!  \sum_{k=1}^{K}  \overline{\nu}_{n,k}\frac{4\pi^2}{\lambda^2}  \Delta \widetilde{\kappa}_{x,k}^2 \cos\!\left (  \overline{\xi}_{n,k} \!+\! \frac{2\pi}{\lambda}\Delta \widetilde{d}_{k,n} \right), \\
    %  &\!\frac{\partial^2 \widetilde{f}_{1,3}}{\partial p_{x,n} \partial p_{y,n}} \!=\! \frac{\partial^2 \widetilde{f}_{1,3}}{\partial p_{y,n} \partial p_{x,n}} \notag \\
    %  &  = \sum_{k=1}^{K}  \overline{\nu}_{n,k}\frac{4\pi^2}{\lambda^2}  \Delta \widetilde{\kappa}_{x,k} \Delta \widetilde{\kappa}_{y,k} \cos\!\left (  \overline{\xi}_{n,k} \!+\! \frac{2\pi}{\lambda}\Delta \widetilde{d}_{k,n} \right), \\
    %   &\!\frac{\partial^2 \widetilde{f}_{1,3}}{\partial p_{y,n}^2} \!=\! \sum_{k=1}^{K}  \overline{\nu}_{n,k}\frac{4\pi^2}{\lambda^2}  \Delta \widetilde{\kappa}_{y,k}^2 \cos\!\left (  \overline{\xi}_{n,k} \!+\! \frac{2\pi}{\lambda}\Delta \widetilde{d}_{k,n}  \right)

By substituting the above results to (\ref{Gradient}) and (\ref{Hessian}), the corresponding gradient and the Hessian matrix of $\widetilde{f}_1$ will be obtained. This finishes the derivations.
\end{appendices}

\bibliographystyle{IEEEtran}
\bibliography{references2}{}

\end{document}